\documentclass[%
 reprint,
superscriptaddress,
nofootinbib,
 amsmath,amssymb,
 onecolumn,
prd,
]{revtex4-2}

\usepackage{graphicx}
\usepackage{dcolumn}
\usepackage{bm}
\usepackage[colorlinks]{hyperref}

\usepackage{color}
\usepackage[caption=false]{subfig}
\def\beq{\begin{eqnarray}}
\def\eeq{\end{eqnarray}}

\usepackage{xspace}
\usepackage{amsmath}

\let\vec\mathbf

\newcommand{\Mpch}{h^{-1}\mathrm{Mpc}}
\newcommand{\hMpc}{h\,\mathrm{Mpc}^{-1}}

\newcommand{\delD}[1]{(2\pi)^3\delta_D\left({#1}\right)}

\newcommand{\fkp}{\mathrm{FKP}}

\newcommand{\av}[1]{\left\langle{#1}\right\rangle} 
\newcommand{\avEst}[1]{\mathbb{E}\left[#1\right]}

\newcommand{\vk}{\vec k}

\newcommand{\vp}{\vec p}
\newcommand{\vd}{\vec d}
\newcommand{\vq}{\vec q}
\newcommand{\vx}{\vec x}
\newcommand{\vy}{\vec y}
\renewcommand{\vr}{\vec r}
\newcommand{\vm}{\vec m}
\newcommand{\vn}{\vec n}
\newcommand{\Hi}{\mathsf{H}^{-1}}
\newcommand{\C}{\mathsf{C}}
\newcommand{\B}{\mathsf{B}}
\newcommand{\Q}{\mathsf{Q}}
\newcommand{\Sig}{\mathsf{S}}
\newcommand{\N}{\mathsf{N}}
\newcommand{\Ci}{\mathsf{C}^{-1}}
\newcommand{\tCi}{\tilde{\mathsf{C}}^{-1}}

\newcommand{\ft}[1]{\mathcal{F}\left[{#1}\right]}
\newcommand{\ift}[1]{\mathcal{F}^{-1}\left[{#1}\right]}
\newcommand{\fid}{\mathrm{fid}}
\newcommand{\Tr}[1]{\operatorname{Tr}\left[{#1}\right]}

\definecolor{darkgreen}{RGB}{0,120,0}

\definecolor{brown}{RGB}{120,60,0}
\newcommand{\resub}[1]{#1}

\begin{document}


\title{Cosmology Without Windows: \\ \small Quadratic Estimators for the Galaxy Power Spectrum}

\author{Oliver H.\,E. Philcox}
\email{ohep2@cantab.ac.uk}
\affiliation{Department of Astrophysical Sciences, Princeton University,\\ Princeton, NJ 08540, USA}%
\affiliation{School of Natural Sciences, Institute for Advanced Study, 1 Einstein Drive,\\ Princeton, NJ 08540, USA}


\begin{abstract}
Conventional algorithms for galaxy power spectrum estimation measure the true spectrum convolved with a survey window function, which, for parameter inference, must be compared with a similarly convolved theory model. In this work, we directly estimate the unwindowed power spectrum multipoles using quadratic estimators akin to those introduced in the late 1990s. Under Gaussian assumptions, these are optimal and free from the leading-order effects of pixellization and non-Poissonian shot-noise. They may be straightforwardly computed given the survey data-set and a suite of simulations of known cosmology. We implement the pixel-based maximum-likelihood estimator and a simplification based on the FKP weighting scheme, both of which can be computed via FFTs and conjugate gradient descent methods. Furthermore, the estimators allow direct computation of spectrum coefficients in an arbitrary linear compression scheme, without needing to first bin the statistic. Applying the technique to a subset of the BOSS DR12 galaxies, we find that the pixel-based quadratic estimators give statistically consistent power spectra, compressed coefficients, and cosmological parameters to those obtained with the usual windowed approaches. Due to the sample's low number density and compact window function, the optimal weighting scheme gives little improvement over the simplified form; this may change for dense surveys or those focusing on primordial non-Gaussianity. The technique is shown to be efficient and robust, and shows significant potential for measuring the windowless power spectrum and bispectrum in the presence of weak non-Gaussianity.
\end{abstract}

\maketitle

\section{Introduction}
The fundamental building block of spectroscopic surveys is the two-point correlator of galaxy positions in real- or Fourier- space. Under the assumption of Gaussianity, this fully specifies the deterministic part of the density field; our Universe may be considered Gaussian on large scales, thus this still encapsulates much of the available information. Measuring such a quantity, be it the two-point correlation function or the power spectrum, has thus become a central tenet of cosmological research throughout the past few decades.

Almost all current surveys \citep[e.g.][]{2017MNRAS.466.2242B,2020MNRAS.498.2492G} have opted to measure the galaxy power spectrum, $P(\vk)$, using the `FKP' approach, first proposed in Ref.\,\citep{1994ApJ...426...23F} and generalized to anisotropic spectra in Refs.\,\citep{2003ApJ...595..577Y,2006PASJ...58...93Y,2015MNRAS.453L..11B,2017JCAP...07..002H}, as well as lightcone evolution in Ref.\,\citep{2017JCAP...04..029S}. This estimates the galaxy power spectrum by performing a Fourier transform on the weighted (galaxy - random) field, where the randoms are a set of artificial galaxies that are unclustered, yet encapsulate the survey geometry and selection function. A weight is applied to each galaxy individually, equal to $1/\left[1+\bar{n}(\vr)P_\fkp\right]$, where $\bar{n}(\vr)$ is the unclustered number density and $P_\fkp$ some constant. As shown in Ref.\,\citep{1998ApJ...499..555T}, this is the minimum variance solution \resub{for Gaussian fluctuations} on small scales, where the spectrum is shot-noise dominated. On large scales however, it is suboptimal, since it (a) does not account for the variations of $P(\vk)$ with respect to scale (assuming $P(\vk)\approx P_\fkp$), and (b) ignores correlations \textit{between} specific pixels, since the weighting is not applied in a pairwise fashion. Furthermore, as is well known, the FKP method estimates the true underlying galaxy power spectrum convolved with the square of the window function \resub{rather than the true spectrum itself}.

In the late 1990s, a slew of work was devoted to avoiding the pitfalls of the FKP estimator, using a more general approach; the \textit{quadratic estimators}. Developed originally for analysis of the Cosmic Microwave Background (CMB) \citep{1997PhRvD..55.5895T,1997ApJ...480...22T,1998PhRvD..57.2117B,1999PhRvD..59b7302B,1999ApJ...510..551O}, these can be similarly applied to galaxy surveys \citep{1998ApJ...499..555T,2005astro.ph..3603H,2005astro.ph..3604H}, and allow one to compute \resub{\textit{unwindowed}} estimates of the underlying power spectra. At heart, a quadratic estimator is simply a quadratic function applied to the data-set, allowing for each pair of particles (or pixels) to be differently weighted. It can be shown that the \textit{optimal} power spectrum estimator is a member of this class (under Gaussian assumptions), thus its application would be expected to give the smallest errors on cosmological parameters. The estimators can be further tuned to weight galaxies also according to their physical properties \citep{2015MNRAS.454.1266S,2016MNRAS.457.4285S}, and used to place competitive constraints on primordial non-Gaussianity \citep{2019JCAP...09..010C}.

If quadratic estimators are so useful, why then have they not been adopted by recent surveys? Indeed, they were used extensively at the start of the twenty-first century \citep[e.g.,][]{2000MNRAS.317L..23H,2002MNRAS.335..887T,2002ApJ...571..191T,2004ApJ...606..702T}, but have since fallen out of vogue. One reason for this lies in computational cost; the optimal estimator is difficult to implement since it requires knowledge of the inverse covariance between any two galaxy positions (or pixels). Current surveys contain $\mathcal{O}(10^6)$ objects, thus a brute-force computation of this inverse is infeasible. Secondly, the FKP scheme is often found to perform well in practice. For a survey containing both uniform number density, $\bar{n}$, and narrow bins, the FKP weights become optimal, just as on small scales. The gains from the optimal approach are expected to be most relevant on scales where the window function has significant power; given that these are usually dominated by cosmic variance, their improved measurement is not expected to have great effect on cosmological analyses, unless one is searching for signatures only visible at low-$k$ (e.g., primordial non-Gaussianity).\footnote{See Ref.\,\citep{2004MNRAS.349..603E} for an extensive comparison of quadratic and FKP-type (psuedo-$C_\ell$) estimators in the CMB context.} Finally, the issue of \resub{windowing} is not usually significant, since our theory-models can be straightforwardly window-convolved \citep[e.g.,][]{2017MNRAS.466.2242B}. That said, in the current era of `precision cosmology', our goal is to squeeze out every last drop of cosmological information, thus any gains, even if small, are of great importance.\footnote{For another notable work, see Ref.\,\citep{2017JCAP...12..009S}, which estimates the \textit{initial} power spectrum via a Bayesian approach.} \resub{Furthermore, the ability to estimate unwindowed spectra will be of great use for higher-order statistics, whereupon window convolution is highly non-trivial.}

In this work, we implement a variant of the quadratic power spectrum estimators of old, considering both the optimal approach and a related scheme which allows for easier computation, akin to the FKP formalism. We provide a pedagogical discussion of their derivation and application, providing forms which can be efficiently applied, bypassing some of the usual computational caveats (following similar approaches to that done for the CMB in Refs.\,\citep{1999ApJ...510..551O,2011MNRAS.417....2S}). The scheme can be similarly used to measure coefficients of the power spectrum under some linear compression scheme, as in Ref.\,\citep{2021PhRvD.103d3508P}, and the accompanying subtleties of this are discussed. By formulating our estimators as a difference between quantities measured in the data and simulations, we are free from the leading-order effects of \resub{pixellization}, binning, non-Poissonian shot-noise, and fingers-of-God, and obtain power spectrum estimates that are not \resub{convolved with} the window function, \resub{with the integral constraint isolated to the first $k$-bin.} Indeed, we demonstrate that the resulting constraints on cosmological parameters from BOSS data analyzed with the quadratic estimators are statistically compatible with those obtained from the usual windowed approach. 

Whilst we do not expect significant gains from applying the optimal estimator to BOSS, this is a strongly survey-dependent statement. For dense small-volume data-sets such as the DESI Bright Galaxy Survey \citep{2016arXiv161100036D}, the improvements will be of greater importance, and the situation is similar for analyses constraining $f_\mathrm{NL}$ (demonstrated in Ref.\,\citep{2019JCAP...09..010C} for the eBOSS quasar sample). The formalism additionally extends beyond the Gaussian limit and can be used to implement optimal estimators for the two- and three-point statistics in the presence of weak non-Gaussianity; our work thus provides a proof-of-concept for such applications. Computing bispectrum estimates free from the survey window function would be a significant achievement, given the inherent difficulty of performing window-convolution in this case \citep[e.g.,][]{2017MNRAS.465.1757G}.




The structure of this paper is as follows. We begin by presenting a pedagogical introduction to the quadratic power spectrum estimator in Sec.\,\ref{sec: qe-theory}, before we discuss its practical implementation in Sec.\,\ref{sec: qe-implementation}. Sec.\,\ref{sec: subspace-qe} details the modifications required for the estimators to measure compressed subspace coefficients, following which we present a demonstration of the approach in Sec.\,\ref{sec: boss-analysis}, computing both spectra and corresponding cosmological parameter constraints. We conclude in Sec.\,\ref{sec: conclusion}, with appendices \ref{appen: qe-properties}\,-\,\ref{appen: pix-options}, providing useful mathematical results, consistency tests and a discussion of the optimal estimators for the power spectrum beyond the Gaussian limit. 
Given the technical nature of this work, we provide a roadmap for two types of readers: those who are principally interested in the quadratic estimator phenomenology and application to galaxy surveys are advised to read Secs.\,\ref{sec: qe-theory}\,-\,\ref{sec: subspace-qe}, whilst those concerned principally with its efficacy when applied to BOSS data (and beyond) should skip directly to Sec.\,\ref{sec: boss-analysis}.

\section{Power Spectrum Estimators}\label{sec: qe-theory}
We begin with a discussion of quadratic power spectrum estimators, discussing both their general and optimal form, before specializing to their application in spectroscopic surveys. Analogous estimators for the coefficients of the power spectrum following some linear compression scheme will be discussed in Sec.\,\ref{sec: subspace-qe}. Several of the derivations below follow Refs.\,\,\cite{1998ApJ...499..555T,1998PhRvD..57.2117B,1999ApJ...510..551O,2004ApJ...606..702T}, but we recapitulate them for clarity.

\subsection{General Quadratic Estimators}\label{subsec: qe-opt-estimators}
Consider a vector $\vd$ of observations, e.g., the pixellized overdensity for a galaxy survey in $N_\mathrm{pix}$ cells. We model this as the sum of two contributions; a signal $\vm$ and a noise component $\vn$, satisfying $\av{\vm\vm^T}=\Sig$ and $\av{\vn\vn^T} = \N$ respectively, for signal and noise covariances $\Sig$ and $\N$, \resub{where $\av{...}$ indicates an average over both the underlying density fields and their Poisson realizations.} Assuming the noise to be uncorrelated with the signal, the covariance of $\vd$ is \resub{given by} $\av{\vd\vd^T} = \C_D=\Sig+\N$. \resub{This can be expressed as a function of band-powers $\vp\equiv \{p_\alpha\}$,\footnote{We use Greek letters $\alpha,\beta,...$ to index band-powers and Latin indices $i,j,...$ to denote components of $\vd$, \textit{i.e.} pixels. \resub{We will later define the band-powers as estimates of some power spectrum multipole integrated over a finite $k$-bin.}} such that $\C_D\equiv\C(\vp^\mathrm{true})$, where $\C(\vp)$ is the covariance defined at using arbitrary set of band-powers $\vp$. If the covariance is a linear function of the band-powers, \textit{i.e.} $\C(\vp)$ is linear in $\vp$,} a general quadratic estimator for $\vp$ is
\beq\label{eq: general-q-alpha-expression}
    \hat{q}^\mathrm{QE}_\alpha = \frac{1}{2}\vd^T\Hi\C_{,\alpha}\Hi\vd \equiv \frac{1}{2}\Tr{\left(\Hi\C_{,\alpha}\Hi\right)\vd\vd^T}
\eeq
where $\C_{,\alpha} = \Sig_{,\alpha} \equiv \frac{\partial \Sig}{\partial p_\alpha}$ is the derivative of the signal with respect to the band-powers and $\mathsf{H}$ is some (positive-definite and symmetric) pixel weighting matrix. Both matrices have dimension $N_\mathrm{pix}\times N_\mathrm{pix}$ and their action is to weight the data by some linear function $\Hi$ (which may or may not be diagonal), then to extract the piece depending on $p_\alpha$ via the derivative term. 

\resub{$\hat{q}_\alpha$} is neither \resub{an} unbiased nor optimal estimator of the band-powers. In expectation, we obtain
\beq
    \avEst{\hat{q}^\mathrm{QE}_\alpha} = \frac{1}{2}\Tr{\Hi\C_{,\alpha}\Hi\C_D} = \frac{1}{2}\Tr{\Hi\C_{,\alpha}\Hi\N} + \frac{1}{2}\sum_{\beta}p^\mathrm{true}_\beta\Tr{\Hi\C_{,\alpha}\Hi\C_{,\beta}},
\eeq
rewriting the data covariance $\C_D$, in terms of the noise covariance and signal derivatives, for true band-powers $\vp^\mathrm{true}$.\footnote{Strictly, this is only true if $\vp$ contains all possible band-powers, \textit{i.e.} it extends over all wavevectors. In practice, truncating to a finite range of wavenumbers is a valid approximation, though, for this reason, one may wish to estimate a slightly larger number of band-powers than used in the final analysis. This point is discussed in more detail in Sec.\,\ref{subsec: sv-qe}.} To debias \eqref{eq: general-q-alpha-expression}, we require both an additive term to subtract the first (noise-induced) contribution to $\hat{q}^\mathrm{QE}_\alpha$ and a multiplicative one to cancel the second trace term. Incorporating these leads to the general quadratic estimator, defined by
\beq\label{eq: general-qe-est}
    \boxed{\hat{p}_\alpha^\mathrm{QE} = p^\mathrm{fid}_\alpha + \sum_\beta F^{-1,\mathrm{QE}}_{\alpha\beta}\left(\hat{q}_\beta^\mathrm{QE}-\bar{q}_\beta^\mathrm{QE}\right),}
\eeq
where we have defined the Fisher and bias terms
\beq\label{eq: general-qe-est-comp}
    F^\mathrm{QE}_{\alpha\beta} &=& \frac{1}{2}\Tr{\Hi \C_{,\alpha}\Hi\C_{,\beta}}\\\nonumber
    \bar{q}^\mathrm{QE}_\alpha &=& 
    \frac{1}{2}\Tr{\Hi\C_{,\alpha}\Hi\C_\fid},
\eeq
again depending on the weighting matrix $\mathsf{H}$. We have further introduced a set of \textit{fiducial} band-powers $\vp^\mathrm{fid}$ \resub{and corresponding covariance $\C_\fid\equiv\C(\vp^\fid)$}; whilst these could alternatively be absorbed into the bias term $\bar{\vec q}$, alongside a redefinition $\bar{q}_\beta \rightarrow \frac{1}{2}\Tr{\Hi\C_{,\beta}\Hi\N}$, we generally avoid this, as it makes the estimators more sensitive to unmeasured $k$-bins, non-Poissonian shot-noise and gridding artefacts. Further discussion of this can be found in Sec.\,\ref{subsec: implement-fid}. The computation of each term in \eqref{eq: general-qe-est} will be elaborated upon in subsequent sections. 

As shown in Appendix \ref{appen: qe-properties}, the general quadratic estimator is unbiased for all choices of symmetric and invertible weighting matrix $\mathsf{H}$ (assuming that the band-powers $\vp$ fully define the signal covariance $\Sig$), such that $\avEst{\hat{p}_\alpha^\mathrm{QE}} = p^\mathrm{true}_\alpha$. This remains valid even when the underlying density field is non-Gaussian, in which case not all information is encoded within the two-point covariance $\C_D$. In the Gaussian limit, the estimator variance is given by
\beq
    \operatorname{cov}\left(\hat{p}^\mathrm{QE}_\alpha, \hat{p}^\mathrm{QE}_\beta\right) = \frac{1}{2}\sum_{\gamma\delta}F^{-1, \mathrm{QE}}_{\alpha\gamma}F^{-1,\mathrm{QE}}_{\beta\delta}\Tr{\Hi\C_{,\gamma}\Hi\C_D\Hi\C_{,\delta}\Hi\C_D},
\eeq
(Appendix \ref{appen: qe-properties}), which depends on our choice of $\mathsf{H}$. Clearly, one must carefully choose the pixel weight matrix to obtain an estimator with low variance; in subsequent sections, we will consider both the minimum variance solution and a simpler approximation.

\subsection{The Maximum-Likelihood Estimator}
Under Gaussian assumptions, the maximum-likelihood (ML) estimator for the band-powers $\vp$ is just a special case of \eqref{eq: general-qe-est}, with $\mathsf{H} = \C_D$. To demonstrate this, we begin by constructing a log-likelihood for the data as
\beq\label{eq: vp-like}
    \mathcal{L}[\vd](\vp) = -2\log L[\vd](\vp) = \vd^T\C^{-1}(\vp)\vd + \operatorname{Tr}\log\C(\vp) + \text{const.}
\eeq
assuming that $\av{\vd}=\vec 0$ (\textit{i.e.} that $\vd$ is mean-subtracted) and that the likelihood is Gaussian (valid if the number of modes is large and the underlying density field is Gaussian). Note that the band-powers enter only through the data covariance $\C(\vp)$, with $\C_D \equiv \C(\vp^\mathrm{true})$, \resub{as before}. 

To obtain an ML estimator for $\vp$, we must extremize \eqref{eq: vp-like}; practically, this is difficult since the band-powers do not enter the log-likelihood linearly, thus we first expand $\mathcal{L}(\vp)$ around some fiducial spectrum $\vp^\fid$;
\beq\label{eq: vp-like-approx}
    \mathcal{L}(\vp^\fid+\delta\vp) \approx \mathcal{L}(\vp^\fid)+\delta\vp^T\nabla_{\vec p}\mathcal{L} + \frac{1}{2}\delta\vp^T\left(\nabla_{\vec p}\nabla_{\vp'}\mathcal{L}\right)\delta\vp',
\eeq
where $\delta \vp = \vp - \vp^\fid$ and the gradients are evaluated at $\vp^\fid$. Assuming $\vp^\fid$ is sufficiently close to the true spectrum (such that $\C_\fid\approx \C_D$), $\delta\vp$ can be estimated by minimizing \eqref{eq: vp-like-approx};
\beq\label{eq: newton-raph-sol}
    \delta\vp = -\left[\nabla_{\vp}\nabla_{\vp'}\mathcal{L}\right]^{-1}\nabla_{\vp'}\mathcal{L},
\eeq
which is \resub{just} a first-order Newton-Raphson estimate. Inserting the Gaussian likelihood and simplifying gives the maximum-likelihood\footnote{Strictly, this is only `maximum-likelihood' if $\vp^\fid$ is equal to the true spectrum. This limit can be achieved by iteration; first define some fiducial spectrum $\vp^{\fid,0}$, which is used in \eqref{eq: newton-raph-sol} to find an updated estimate $\vp^{\fid,1}$ and repeated until convergence is reached. Whilst this is possible (and discussed in Ref.\,\cite{1998PhRvD..57.2117B}) it is time-consuming, and the estimator is no longer strictly quadratic. Furthermore, the degree of suboptimality is quadratic in the difference between $\Ci_D$ and $\Ci_\fid$, and thus usually small.} estimator for a band-power $p_\alpha$;
\beq\label{eq: quadratic-estimator-def}
    \boxed{\hat{p}_\alpha^\mathrm{ML} = p_\alpha^\fid + \frac{1}{2}\sum_\beta F^{-1}_{\alpha\beta}\left(\vd^T\Ci\C_{,\beta}\Ci_\fid\vd-\Tr{\Ci_\fid\C_{,\beta}}\right),}
\eeq
where we have assumed (as is standard practice) that the curvature matrix $\nabla_{\vp}\nabla_{\vp'}\mathcal{L}$ can be replaced by its (realization-averaged) Fisher matrix, defined by
\beq
    F_{\alpha\beta} = \frac{1}{2}\Tr{\Ci_\fid\C_{,\alpha}\Ci_\fid\C_{,\beta}}.
\eeq
\cite{1998ApJ...499..555T,1998PhRvD..57.2117B,1999ApJ...510..551O,2004ApJ...606..702T}.\footnote{See \citep{1999PhRvD..59b7302B} for a CMB analysis that does not make this assumption.} Writing this in the form of \eqref{eq: general-qe-est};
\beq\label{eq: quad-est-split}
    \hat{p}^\mathrm{ML}_\alpha = p_\alpha^\fid + \sum_{\beta} F^{-1}_{\alpha\beta}\left(\hat{q}_\beta - \bar{q}_\beta\right),
\eeq
we see that this is just a special case of the general quadratic estimator, with $\mathsf{H}$ equal to the covariance matrix in the fiducial cosmology $\C_\fid$.

As shown in Appendix \ref{appen: qe-properties}, the ML solution is optimal (in the Cram\'er-Rao sense) if $\C_\fid$ is equal to the true data covariance $\C_D$ (\textit{i.e.} $\vp^\fid = \vp^\mathrm{true}$) and the underlying density field is Gaussian, such that all connected moments of the density field above the two-point function are vanishing. In this case, the covariance of $\hat{p}_\alpha$ is simply the inverse Fisher matrix $F_{\alpha\beta}^{-1}$. 

In reality, the Universe is non-Gaussian except on the largest scales, thus the quadratic estimator is not strictly optimal. In the weakly non-Gaussian limit, one may construct the leading order corrections to the maximum-likelihood estimator for $\vp^\fid$ via an Edgeworth expansion \citep[e.g.,][]{2017arXiv170903452S}. These start cubic in the data, and are discussed in Appendix \ref{appen: non-Gaussian-estimator}. For most galaxy surveys, this is of limited use, since the power spectrum becomes dominated by shot-noise on relatively large-scales, though the approach may be of use for dense surveys such as the DESI Bright Galaxy Survey \citep{2016arXiv161100036D}.

\subsection{Specialization to Spectroscopic Surveys}\label{subsec: qe-spectro}
\subsubsection{Covariance Definitions}
The quadratic estimator \eqref{eq: general-qe-est} and its optimal variant \eqref{eq: quadratic-estimator-def} provide a general framework in which to measure a set of band-powers $\vp$ from a data-set $\vd$ given the signal and noise covariances. For spectroscopic surveys, our data-set is a list of galaxy positions alongside a set of random particles which encode the survey geometry. Here, we define $\vd$ as the pixellized field of (data - randoms), \textit{i.e.}
\beq
    d_i \equiv n_g(\vr_i)-\alpha\,n_r(\vr_i)    
\eeq
where $\vr_i$ is the center of the $i$-th pixel and $n_g$ ($n_r$) is the galaxy (random particle) density defined on a grid following some mass assignment scheme, for example cloud-in-cell or the more nuanced schemes of Ref.\,\cite{1998ApJ...499..555T} (which can additionally be used to null certain troublesome modes, such as those dominated by fingers-of-God effects). The coefficient $\alpha$ is chosen such that $\av{\vd} = 0$, \textit{i.e.} $\alpha = \int d\vr\,n_g(\vr)/\int d\vr\,n_r(\vr)$.\footnote{We may additionally include particle weights in the definition of $\vd$, for example to encapsulate systematics and fiber collisions (but not FKP weights). \resub{This gives $d_i \equiv w_g(\vr_i)n_g(\vr_i) - \alpha\,w_r(\vr_i)n_r(\vr_i)$ where $\alpha = \int d\vr\,w_g(\vr)n_g(\vr)/\int d\vr\,w_r(\vr)n_r(\vr)$.} Such weights are included in the analysis of data in Sec.\,\ref{sec: boss-analysis}.}

The covariance matrix of $\vd$ is given by $\C_{ij} = \C(\vr_i,\vr_j)$,\footnote{For interpretability, we write most quantities in continuous form, noting that any integrals are strictly sums over the pixel grid, and any Fourier transforms are discrete.} where
\beq
    \C(\vr,\vr') &=& \av{[n_g-\alpha\,n_r](\vr)[n_g-\alpha\,n_r](\vr')}\\\nonumber
    &=& n(\vr)n(\vr')\xi(\vr,\vr') + (1+\alpha)n(\vr)\delta_D(\vr-\vr')\\\nonumber
    &\equiv& \Sig(\vr,\vr') + \N(\vr,\vr'),
\eeq
assuming Poisson statistics and writing $\av{n_g(\vr)} = \alpha\av{n_r(\vr)} = n(\vr)$ \resub{(suppressing the dependence of $\C$ on $\vp$ for clarity)}. We have used that the zero-noise part of the galaxy density field can be written $n_g(\vr) = n(\vr)\left[1+\delta(\vr)\right]$, where $\delta(\vr)$ has the correlation function $\xi(\vr,\vr') = \av{\delta(\vr)\delta(\vr')}$, and ignored the effects of the \resub{pixellization} scheme.\footnote{It is possible to self-consistently include this, by noting that any pixellized quantity $\tilde{x}$ is equal to $\tilde{x}\equiv \psi\ast x$, where $x$ is the underlying field, $\ast$ indicates a convolution and $\psi$ is the mass assignment window function. Thus, $\C(\vr,\vr')\rightarrow\left[\psi\ast\C\ast\psi\right](\vr,\vr')$. This is more difficult to evaluate however, since the noise covariance is no longer diagonal, and the field $n(\vr)$ appearing in $\C$ is the \textit{unconvolved} mean field, estimated from $\psi^{-1}\ast\tilde{n}$. This leads to a significant increase in the number of Fourier transforms that need be performed to compute any quantity involving $\C$. Ignoring these effects is a valid assumption in practice, since they are cancelled at leading-order in the difference estimator of \eqref{eq: general-qe-est}.} Writing the correlation function (which, in the general case, will be that of \resub{galaxies} in redshift-space) in Fourier-space gives\footnote{We neglect the impact of redshift evolution on the galaxy power spectrum, which is valid if the radial extent of the sample is relatively small. One may alternatively include this at linear order, writing $P(k, z) \approx P(k, z_\mathrm{eff})\times\left(D(z)/D(z_\mathrm{eff})\right)^2$, leading to two additional (spatially varying) factors of $D(z)/D(z_\mathrm{eff})$ being inserted into the covariance, giving a slightly modified weight. \resub{Technically, this is true only in the absence of RSD, since $f(z)$ evolves differently in redshift to $D(z)$. Full treatment of this is beyond the scope of this work, though is important for $f_\mathrm{NL}$ based studies such as Ref.\,\citep{2019JCAP...09..010C}.}}
\beq\label{eq: C-def}
    \C(\vr,\vr') = n(\vr)n(\vr')\int_{\vk}e^{i\vk\cdot(\vr-\vr')}\sum_\ell P_\ell(k)L_\ell(\hat{\vk}\cdot\hat{\vr}') + (1+\alpha)n(\vr)\delta_D(\vr-\vr'),
\eeq
denoting $\int_{\vk}\equiv (2\pi)^{-3}\int d\vk$ and expanding in multipoles via the Legendre polynomial $L_\ell$, as in the Yamamoto formalism for redshift-space distortions \citep{2006PASJ...58...93Y}.\footnote{A more accurate case would be to expand $\xi(\vr,\vr')$ in the angle between $(\vr-\vr')$ and $(\vr+\vr')/2$, as in Ref.\,\citep{2021arXiv210208384P}. The Yamamoto approximation assumes that the line-of-sight \resub{can be approximated as} $\vr'$, which goes beyond the flat-sky approximation. \resub{For the most general treatment, though one beyond the scope of this work, one would instead allow the power spectrum to depend on \textit{two} lines of sight.}} We may similarly define the band-power derivatives, assuming $P_\ell(k) = \sum_a \Theta_a(\vk)P_\ell^a$ for some (thin) $k$-bins with bin weights $\Theta^a(\vk)$ which are unity if $\vk$ is in $|\vk|$-bin $a$ and zero else;
\beq\label{eq: C-alpha-def}
    \C_{,\alpha}(\vr,\vr') = n(\vr)n(\vr')\int_{\vk}e^{i\vk\cdot(\vr-\vr')}\Theta_a(\vk)L_\ell(\hat{\vk}\cdot\hat{\vr}'),
\eeq
where the index $\alpha$ refers to the $\{a,\ell\}$ pair. The above formulae allow computation of the band-powers $\vp\equiv \{P_\ell^a\}$ via \eqref{eq: general-qe-est} or \eqref{eq: quadratic-estimator-def}, in particular, allowing for the implementation of a close-to-optimal estimator, by applying the $\Ci_\fid$ weights to the (data - random) pixel grid. 

\subsubsection{Relation to Standard Estimators}
Before continuing, we pause briefly to compare the ML estimator of \eqref{eq: quadratic-estimator-def} with the windowed FKP estimator of Ref.\,\citep{1994ApJ...426...23F} conventionally used in the literature (or its more modern variants \citep{2015MNRAS.453L..11B,2017JCAP...07..002H}). Although the ML approach requires significantly more computational expense than a simple FKP estimate (mainly due to the necessity to invert the \resub{fiducial} covariance $\C_\fid$), it has two key benefits: (1) it produces minimum variance error bars on the power spectrum (and hence derived parameters), with a particular improvement seen on large scales (the FKP weights are optimal on short-scales \citep{1998ApJ...499..555T}); (2) by including the noise model and survey geometry in the covariances $\C(\vp)$, the estimator recovers the \textit{true} unwindowed power spectrum. Furthermore, if one assumes that the shot-noise is treated correctly in the fiducial power spectrum, the band-power estimates will also be shot-noise free; in general, we will leave a free shot-noise in the eventual parameter analysis, to account for its possible non-Poissonian nature (which may be different in data and simulations).

Our approach requires a fiducial (unwindowed) power spectrum, and, for practical application, a set of simulated realizations, with the same geometry and noise properties as the data, (see Sec.\,\ref{sec: qe-implementation}). Practically, we may use a suite of $N$-body simulations for this purpose. Though the unwindowed galaxy power-spectra are not usually known \textit{a priori}, one can fit the mean of the windowed FKP mock spectra with a \resub{window}-convolved model such as one-loop \resub{Effective Field Theory (hereafter EFT)} \citep{2020JCAP...05..042I} (given that the cosmology is already known), and use the pre-convolved best-fit to set the fiducial power spectrum. \resub{To avoid biasing the estimated power spectra, we require the theory model to be accurate, such that the model power spectrum is close to the truth; this just requires the $\chi^2$ of the fit to be small. There is no requirement for the fiducial model to match that used to analyze the data however, thus we do not need to recompute the statistic when comparing different theory prescriptions.} 
An alternative approach would be to \textit{deconvolve} \resub{the window from the simulated power spectra to use as a fiducial model}, though this is difficult given the finite $k$-range of the observed spectra. \resub{We further note that it is possible to remove dependence on the fiducial power spectrum entirely (except in the weights), as discussed below \eqref{eq: qe-no-fid}, though we retain it to reduce some higher-order effects such as discretization bias.}

\subsubsection{FKP Weights}

As mentioned above, the ML estimator requires the inverse covariance matrix, $\Ci_\fid$, which, due to its high dimensionality, is difficult and time-consuming, to compute. For this reason, we consider an alternative quadratic estimator in the form of \eqref{eq: general-qe-est}, with weights based on the FKP approach. This uses the weighting matrix
\beq\label{eq: H-fkp-def}
    \mathsf{H}_\fkp(\vr,\vr') = n(\vr)n(\vr')P_\fkp\delta_D(\vr-\vr') + n(\vr)\delta_D(\vr-\vr'),
\eeq
which may be seen as the small-scale limit of $\C$ with $P(\vk)\rightarrow P_\fkp\sim 10^4h^{-3}\mathrm{Mpc}^3$ \citep{1998ApJ...499..555T}, additionally setting $\alpha = 0$. Equivalently, this is a simplified form of $\C$, assuming the window function to be compact, \resub{such that its characteristic width $L$ satisfies $kL\gg 1$ for all scales of interest $k$}. This is trivially inverted;
\beq\label{eq: H-inv-fkp-def}
    \mathsf{H}_\fkp^{-1}(\vr,\vr') = \frac{1}{n(\vr)\left[1+n(\vr)P_\fkp\right]}\delta_D(\vr-\vr').
\eeq
Inserting this into the general quadratic estimator \eqref{eq: general-qe-est}, by construction, gives an unbiased estimator of the unwindowed galaxy power spectrum, which we expect to approach the optimal solution on small-scales. Note that this is \textit{not} identical to the conventional FKP scheme of Ref.\,\cite{1994ApJ...426...23F}, which assigns weights $w_\fkp(\vr) = 1/\left(1+\bar{n}(\vr)P_\fkp\right)$ to galaxies and random particles, where $\bar{n}(\vr)$ is some (spatially-varying) mean density. Our formalism applies the weight to the grid directly rather than the particles, and normalizes by $n(\vr)$ rather than some survey-averaged quantity; this is necessary to ensure that we recover the \textit{unwindowed} power spectrum estimates, unlike in Ref.\,\cite{1994ApJ...426...23F}. In the below, we will compare band-power estimates computed using $\mathsf{H}_\fkp$ to those with the full fiducial covariance $\C_\fid$. We note that the quadratic scheme with FKP weights is similar to that proposed in Ref.\,\citep{2019JCAP...09..010C} for the measurement of quasar spectra, though their formalism still required window-function convolution.

\section{Implementation}\label{sec: qe-implementation}
To apply the power spectrum estimators of Sec.\,\ref{sec: qe-theory}, we perform the following procedure:
\begin{enumerate}
    \item Given a set of simulations with known cosmology, fit the mean window-convolved power spectrum to the one-loop EFT model of Ref.\,\cite{2020JCAP...05..042I} to determine the best-fit nuisance parameters. The unconvolved theory model (subtracting Poissonian shot-noise) gives a smooth model for the fiducial $\vp^\fid$ spectra;
    \item Paint the data and randoms to a regular Cartesian grid to define the data-vector $\vd$;
    \item For each ($|\vk|$ and multipole) bin $\alpha$, compute the quadratic estimator terms $\vd^T \Hi\C_{,\alpha}\Hi \vd$ with $\mathsf{H} = \mathsf{H}_\fkp$ \eqref{eq: H-fkp-def} or $\mathsf{H} = \C_\fid$ \eqref{eq: C-def}. For the latter case, covariances are computed using the (smooth) fiducial power spectra;
    \item Compute the bias term and Fisher matrix, as defined in \eqref{eq: quad-est-split} and \eqref{eq: general-qe-est-comp} for the ML and FKP estimators. In both cases, this is done via Monte Carlo methods, using a set of simulations which satisfy $\av{\vd \vd^T} = \mathsf{H}$ (either the simulations used to define the fiducial cosmology for the ML estimators, or rescaled subsets of the random particles for the FKP case);
    \item Compute the fiducial power spectrum $\vp^\fid$ in the relevant bins and multipoles from the smooth model;
    \item Accumulate the estimators \eqref{eq: general-qe-est} or \eqref{eq: quadratic-estimator-def}, using the $\mathsf{H}_\mathrm{FKP}^{-1}$ or $\Ci_\fid$ pixel weight matrices.
\end{enumerate}

Below, we give practical details concerning the implementation of each of the above steps. All computations are performed using custom Python code making use of the \texttt{pyfftw} library, with some elements adapted from the \texttt{pylians} and \texttt{nbodykit} \cite{2018AJ....156..160H} packages.

\subsection{Computation of $\C_\fid$ and $\C_{,\alpha}$}\label{subsec: implement-C-Ca}
To implement the ML quadratic estimators, we must be able to apply the $\C_\fid$ and $\C_{,\alpha}$ matrices to the pixellized galaxy field.\footnote{Whilst $\C_\fid$ does not appear directly in \eqref{eq: quadratic-estimator-def}, it is required to efficiently compute the action of $\Ci_\fid$.} Assuming the density field to have a total of $N_\mathrm{pix}\sim 10^7$ pixels, the matrices have dimension $N_\mathrm{pix}^2$, thus their explicit computation and storage is infeasible. Instead, we consider their action on a general pixellized field $x(\vr)$, starting from the definition \eqref{eq: C-def};\footnote{\resub{We ignore the dependence of $\C$ on $\vp$ througout this section for clarity.}}
\beq
    \C[x](\vr) &\equiv& \int d{\vr'}\,\C(\vr,\vr')x(\vr')\\\nonumber
    &=& n(\vr)\int_{\vk}e^{i\vk\cdot\vr}\int d\vr'\,e^{-i\vk\cdot\vr'}n(\vr')x(\vr')\sum_\ell P_\ell(k)L_\ell(\hat{\vk}\cdot\hat{\vr}') + (1+\alpha)n(\vr)x(\vr).
\eeq
Here $n(\vr)$ is the unclustered galaxy density (depending on the survey selection function and mask), which we set equal to the pixellized density field of random particles, and $\{P_\ell(k)\}$ are the fiducial (unwindowed) power spectrum multipoles. In the flat-sky limit, $L_\ell(\hat{\vk}\cdot\hat{\vr}')\rightarrow L_\ell(\hat{\vk}\cdot\hat{\vec{z}})$ (\textit{i.e.} we have a uniform line-of-sight $\hat{\vec{z}}$ across the survey), this is straightforward to implement; the signal covariance term is $n(\vr)\ift{P(\vk)\ft{nx}(\vk)}(\vr)$ for forward (inverse) Fourier operator $\mathcal{F}$ ($\mathcal{F}^{-1}$). In the Yamamoto case, we can expand $L_\ell$ via the spherical harmonic addition theorem \citep[Eq.\,14.30.9]{nist_dlmf}, yielding
\beq\label{eq: applyC}
    \C[x](\vr) &=& n(\vr)\sum_{\ell}\frac{4\pi}{2\ell+1}\sum_{m=-\ell}^\ell \int_{\vk}P_\ell(k)e^{i\vk\cdot\vr}Y_{\ell m}^*(\hat{\vk})\int d\vr'\,Y_{\ell m}(\vr')n(\vr')x(\vr')e^{-i\vk\cdot\vr'} + (1+\alpha)n(\vr)x(\vr)\\\nonumber
    &=& n(\vr)\sum_{\ell}\frac{4\pi}{2\ell+1}\sum_{m=-\ell}^\ell\ift{P_\ell(k)Y^*_{\ell m}(\hat{\vk})\ft{Y_{\ell m}nx}(\vk)}(\vr) + (1+\alpha)n(\vr)x(\vr).
\eeq
In practice, this is computed with $\tfrac{1}{2}(1+\ell_\mathrm{max})(2+\ell_\mathrm{max})$ real-to-Fourier FFTs using the real form of the spherical harmonics, as in Ref.\,\cite{2017JCAP...07..002H}. This assumes the fiducial spectra are non-zero for all even $\ell$ up to $\ell_\mathrm{max}$. A similar expression holds for the application of $\C_{,\alpha}$ to a generic map $x$ starting from \eqref{eq: C-alpha-def};
\beq\label{eq: applyC-alpha}
    \C_{,\alpha}[x](\vr) = n(\vr)\frac{4\pi}{2\ell+1}\sum_{m=-\ell}^\ell\ift{\Theta_a(\vk)Y^*_{\ell m}(\hat{\vk}) \ft{Y_{\ell m}nx}(\vk)}(\vr)
\eeq
requiring only $(2\ell+1)$ FFTs. Whilst it is somewhat simpler to consider the action of $\C_{,\alpha}$ on a \textit{pair} of fields (which requires two real-to-complex Fourier transforms and a Fourier-space summation), the action of $\C_{,\alpha}$ on a \textit{single} vector is useful when we wish to apply a number of operations in series, \textit{i.e.} $\C_{,\alpha}$ then $\Ci_\fid$.

\subsection{Computation of $\Ci_\fid$ and $\mathsf{H}_\fkp^{-1}$}
Evaluation of the ML estimator requires the inverse-fiducial-covariance weighted data-vector, $\Ci_\fid \vd$. Given that the covariance $\C_\fid$ is too large to be stored and manipulated, we cannot compute $\Ci_\fid$ directly, thus we instead opt to use preconditioned conjugate-gradient descent (CGD) methods, as in Ref.\,\cite{1999ApJ...510..551O}. Effectively, this numerically solves the equation $\Ci_\fid \vx = \vy$ by finding $\vy$ such that $\tCi\C_\fid\vy = \tCi\vx$, where $\tCi$ is some preconditioner matrix. This requires repeated application of $\C_\fid$ to data, possible via \eqref{eq: applyC}. Convergence is expedited by sensible choice of $\tilde{\mathsf{C}}$; we require it to be easily inverted and close to $\C_\fid$, in the sense that $\tCi\C_\fid = \mathsf{I} + \mathsf{R}$ where all the eigenvalues of $\mathsf{R}$ are less than unity. Here we consider the simple choice;
\beq
    \tilde{\mathsf{C}}(\vr,\vr') = \mathsf{H}_\fkp(\vr,\vr') = n(\vr)\left[1+n(\vr)P_\fkp\right]\delta_D(\vr-\vr')
\eeq
with $P_\fkp = 10^4h^{-3}\mathrm{Mpc}^3$ (as in Ref.\,\citep{2015MNRAS.453L..11B}), which is expected to reasonably well-approximate $\C_\fid$. Given this form, we usually find the matrix inversion algorithm to converge (easily tested by computing $\vec x-\C_\fid\tilde{\vec y}$) in around $N_\mathrm{it} \approx 50$ iterations. This is a computationally intensive step, requiring $\tfrac{1}{2}N_\mathrm{it}(1+\ell_\mathrm{max})(2+\ell_\mathrm{max})$ FFTs.

The action of $\mathsf{H}_\fkp$ on an arbitrary map $x$ is required both for the CGD step of the ML estimator and the unwindowed FKP approach. Due to the Dirac function, this is straightforward;
\beq\label{eq: applyHinv}
    \mathsf{H}^{-1}_\fkp[x](\vr) = \frac{x(\vr)}{n(\vr)\left[1+n(\vr)P_\fkp\right]},
\eeq
requiring a simple division in configuration-space.\footnote{Note that one requires $n(\vr)>0$ for the inverse map $\mathsf{H}^{-1}_\fkp[x]$ to be well-defined. For an infinitely-large random catalog, violation of this condition occurs outside the survey mask, but may be ignored since $x(\vr)$ will always be zero in these regions. For a finite number of randoms, it is possible for pixels allowed by the survey grid to contain no random particles. To avoid infinities due to the inversion (which will be present also in the full $\Ci_\fid[x]$ map), we replace such pixels by the average of their 26 nearest neighbours. Reducing the size of the random catalog from $50\times$ to $10\times$ shows this to be of little importance however, even with the finest choice of cell-size.}

\subsection{Bias Term and Fisher Matrix}\label{subsec: implement-bias-fish}

Using the above definitions, we can apply the weighting matrix $\Hi\C_{,\alpha}\Hi$ to the data-vector to find $\hat{q}_\alpha$, first computing $\vec e\equiv \Hi\vd$ and $\C_{,\alpha}\vec e$, then taking the trace $\Tr{\vec e^T\C_{,\alpha}\vec e}$; simply a summed product of two density fields. Below, we consider how to compute the bias and Fisher terms, which proceed slightly differently for the two choices of weight matrix.

\subsubsection{ML Estimators}
The ML bias and Fisher terms are given by $\bar{q}_\alpha = \frac{1}{2}\Tr{\Ci_\fid\C_{,\alpha}}$ and $F_{\alpha\beta} = \frac{1}{2}\Tr{\Ci_\fid\C_{,\alpha}\Ci_\fid\C_{,\beta}}$. Since the required matrices are dense and we do not have access to their full form, we compute these terms via Monte Carlo methods, as in Refs.\,\cite{1999ApJ...510..551O,2011MNRAS.417....2S}, first writing
\beq\label{eq: bias-fish-as-mc}
    2\bar{q}_\alpha = \av{\vm^T\Ci_\fid\C_{,\alpha}\Ci_\fid\vm}, \quad 2F_{\alpha\beta} = \av{\vm^T\Ci_\fid\C_{,\alpha}\Ci_\fid\C_{,\beta}\Ci_\fid\vm}
\eeq
where the random fields $\vm$ satisfy $\av{\vm\vm^T} = \C_\fid$.\footnote{\resub{In fact, we can use any invertible matrix $\mathsf{A}$ to make the Fisher matrix separable, writing $2F_{\alpha\beta} = \av{\vec a^T\Ci_\fid\C_{,\alpha}\Ci_\fid\C_{,\beta}\mathsf{A}^{-1}\vec{a}}$ for $\mathsf{A} = \av{\vec a\vec a^T}$. We do not apply this here however, given that we use Monte Carlo simulations already to define the bias term. }} In practice, we can use $N$-body simulations for these, in particular those used to define the fiducial band-powers $\vp^\fid$, which satisfy the constraint by definition. Computing these traces via Monte Carlo averaging increases the estimator variance by a factor $\sqrt{1+1/N_\mathrm{mc}}$ when $N_\mathrm{mc}$ simulations are used, \textit{i.e.} only $\sim 0.5\%$ for 100 simulations. \resub{For both $\bar{q}_\alpha$ and $F_{\alpha\beta}$, bias will only arise if the equation $\av{\vm\vm^T} = \C(\vp^\fid)$ is not satisfied, \textit{i.e.} if our fiducial model is not accurate. Whilst we technically require $\vp^\fid=\vp^\mathrm{true}$ for optimality, the increase in output parameter variances resulting from an incorrect $N$-body cosmology are expected to be small.} In practice, we compute the contributions via the following algorithm:
\begin{itemize}
    \item For each simulation $\vm$:
    \begin{itemize}
        \item Compute $\Ci_\fid\vm$ via preconditioned CGD using \eqref{eq: applyC};
        \item For each band-power $\alpha$:
        \begin{itemize}
            \item Compute $\vy_{\alpha} = \C_{,\alpha}[\Ci_\fid\vm]$ via \eqref{eq: applyC-alpha};
            \item Accumulate the bias contribution $(\Ci_\fid\vm)^T\vy_{\alpha}$;
            \item Compute $\Ci_\fid[\vy_\alpha]$ via CGD;
            \item Accumulate the Fisher matrix contribution $\frac{1}{2}\vy_\beta^T \Ci_\fid[\vy_\alpha]$ from each choice of $\beta$.
        \end{itemize}
    \end{itemize}
\end{itemize}
Note that we recompute $\vy_\beta$ for each choice of $\alpha$ rather than first storing all $N_\mathrm{bins}$ vectors. This is done to avoid excessive memory requirements, but slows down the estimator. In total, we require $N_\mathrm{mc}(N_\mathrm{bins}+1)$ matrix inversions to perform the above algorithm, and must apply either the $\C_\fid$ or $\C_{,\alpha}$ operator $N_\mathrm{mc}\left[N_\mathrm{it}+ N_\mathrm{bins}(N_\mathrm{it}+N_\mathrm{bins})\right]$ times, where $N_\mathrm{it}$ is the number of CGD steps. A faster approach, and one advocated for by Ref.\,\cite{1999ApJ...510..551O} would be to instead use an approximate form of the Fisher matrix, such as that obtained using the preconditioner matrix $\tCi$ instead of $\Ci_\fid$. If the estimator is iterated until convergence (updating $\vp^\fid$ at each step), the final estimate does not depend on the exact form of $F_{\alpha\beta}$, justifying this approach. In our context, it is expensive to iterate the algorithm, since one would require a new set of simulations for each $\vp^\fid$ update, and these are expensive to compute, since we require them to reproduce the survey geometry and lightcone structure. When \resub{we do not recalibrate the simulation cosmology}, as here, the full Fisher matrix is required. An additional approach would be set $F_{\alpha\beta}$ equal to the covariance of the $\vm^T\Ci_\fid\C_{,\alpha}\Ci_\fid\vm$ coefficients (which are already used to define the bias vector), noting that this relation is exact for a fully optimal estimator (cf.\,Appendix \ref{appen: qe-properties}). This is also infeasible however, since we require a large number of simulations (\textit{i.e.} $N_\mathrm{mc} \gg N_\mathrm{bins}$) to ensure that the Fisher matrix can be easily inverted. Furthermore, this will include additional higher order contributions from the non-Gaussian four-point function, \textit{i.e.} $\av{d_id_jd_kd_l}$, which lead to a non-trivial bias.

\subsubsection{FKP Estimators}
To compute the bias and Fisher matrices for the quadratic estimator using the FKP weighting \eqref{eq: H-fkp-def}, a similar procedure is possible. Analogously to the ML case, we write
\beq
    \bar{q}_\alpha^\mathrm{QE} &=& \frac{1}{2}\Tr{\Hi\C_{,\alpha}\Hi\C_\fid} = \frac{1}{2}\av{\vm^T\Hi\C_{,\alpha}\Hi\vm}\\\nonumber
    F_{\alpha\beta}^\mathrm{QE} &=& \frac{1}{2}\Tr{\Hi\C_{,\alpha}\Hi\C_{,\beta}} = \frac{1}{2}\av{\tilde{\vm}^T\Hi\C_{,\alpha}\Hi\C_{,\beta}\Hi\tilde{\vm}},
\eeq
where the random fields $\vm$ and $\tilde{\vm}$ satisfy $\av{\vm\vm^T} = \C_\fid$, $\av{\tilde{\vm}\tilde{\vm}^T} = \mathsf{H}$. As above, we can draw $\vm$ from the set of simulations at fiducial cosmology (thus $\bar{q}_\alpha$ is simply the average of $\hat{q}_\alpha$ over this set of simulations), but the Monte Carlo simulations used to define the \resub{FKP} Fisher matrix require a different definition, \resub{since they must have covariance $\mathsf{H}$ rather than $\C_\fid$}. These can be obtained from the set of random particle positions; we split the random catalog into $\approx 1/\alpha$ disjoint pieces (where $\alpha$ is the data-to-random ratio of $\vd$) and compute $\tilde{m}_i$ as the difference between each of these and the remainder of the randoms, just as for $\vd$. Each density field is then rescaled:
\beq
    \tilde{m}(\vr) \rightarrow \tilde{m}(\vr)\times \sqrt{\frac{1+n(\vr)P_\fkp}{1+\alpha}},
\eeq
such that $\av{\tilde{\vm}\tilde{\vm}^T} = \mathsf{H}_\fkp$. Computation of the bias and Fisher matrix then proceeds as above. Due its diagonal form, $\Hi_\fkp$ can be applied much faster than $\Ci_\fid$; however, the process still requires $N_\mathrm{mc}N_\mathrm{bins}^2$ applications of $\C_{,\alpha}$ for the Fisher matrix, and $N_\mathrm{mc}N_\mathrm{bins}$ for the bias (unless intermediate quantities are stored), thus the algorithm is not significantly faster unless $N_\mathrm{bins}\ll N_\mathrm{it}\sim 50$. Note also that \textit{two} sets of simulations are required in this case; one each for the bias and Fisher terms.

\subsection{Fiducial Spectrum}\label{subsec: implement-fid}
The final component of quadratic estimators \eqref{eq: general-qe-est}\,\&\,\eqref{eq: quadratic-estimator-def} is the set of fiducial band-powers $\vp^\fid \equiv \{P_\ell^a\}$, which are related to the continuous power-spectrum model $P(\vk)$ used to define the fiducial covariance $\C_\fid$. We consider two possibilities for their computation: (1) use the smooth fiducial model evaluated at the bin-centers, or (2) average the fiducial model over the true $k$-bins, incorporating the full complexities of binning and pixellization. The latter case proceeds by first computing $P(\vk)$ on the discrete 3D $\vk$-grid, then binning it in the same manner as the estimators \eqref{eq: applyC-alpha}, leading to
\beq\label{eq: p-fid-discrete}
    p_\alpha^{\fid,\mathrm{discrete}} = \frac{4\pi}{\int_\vk\Theta_a(\vk)}\left[\int_{\vk}\sum_L \frac{4\pi}{2L+1}P_L(\vk)Y^*_{\ell m}(\hat{\vk})Y^{}_{LM}(\hat{\vk})\int \frac{d\vec x}{V}Y_{\ell m}(\hat{\vec x})Y^*_{LM}(\hat{\vec x})\right]
\eeq
where the denominator is just the number of modes in the $|\vk|$-bin and the integrals are sums over all pixels (which have total real-space volume $V$). This is simply derived from the usual Yamamoto power spectrum estimator, and, in the limit of infinitely fine pixels and bins, $p_\alpha^{\fid,\mathrm{discrete}} = P_\ell^a$, as expected.

In this work, we instead adopt the former option, evaluating the theory model at the bin centers, \textit{i.e.} $P_\ell^a = P_\ell(k_a)$ where $k_a$ lies at the center of bin $a$. This is of twofold utility: firstly, it ensures that the fiducial spectra match the theory model used in the later parameter inference (which does not conventionally include bin integration); secondly, it removes the leading-order effects of pixellization and gridding in our statistics, \resub{thus reducing bias}. This occurs since our estimator is fundamentally the fiducial spectra plus the difference of $\hat{q}_\alpha$ computed in data and simulations; any effects that are common to both simulations and data will cancel at lowest order. In practice this allows us to use lower resolution grids to compute the spectra than possible with conventional approaches. Furthermore, using such a difference estimator (with a Monte Carlo-calibrated bias term) can additionally reduce difficult-to-model non-linear effects appearing at low-$k$, such as the fingers-of-God (FoG) effect and non-Poissonian shot-noise, provided they are present both in simulations and data. 

One further comment on the fiducial spectrum is required. The FKP-based quadratic estimator discussed above relies on both a fiducial model for $P(\vk)$ both to define $\vp^\fid$ and to compute the bias (via a set of simulated realizations of this cosmology). Given that the standard FKP estimator for the \textit{windowed} power spectrum multipoles \citep[e.g.,][]{2017JCAP...07..002H} does not require a fiducial model, one may ask whether it is strictly necessary for our unwindowed estimators. As expected, the dependence on $\vp^\fid$ may be completely removed by rewriting \eqref{eq: general-qe-est} in the form
\beq\label{eq: qe-no-fid}
    \hat{p}_\alpha^\fkp \rightarrow \sum_\beta F^{-1,\mathrm{QE}}_{\alpha\beta}\Tr{(\vd\vd^T-\N)\Hi\C_{,\alpha}\Hi},
\eeq
which involves only a redefinition of the bias (and is the form given in Ref.\,\citep{2004ApJ...606..702T}). Here, the bias term depends only on the noise covariance $\N$ (assuming that the signal covariance is fully characterized by the set of all linear band-powers), which, for FKP weights, could be estimated using subsets of the random catalog, just as for $F_{\alpha\beta}^\mathrm{QE}$ above, or via simple Fourier transforms of $n(\vr)$, as detailed in Appendix \ref{appen: fkp-simplif}. \eqref{eq: qe-no-fid} applies also to the ML estimator, though a fiducial model is still required to compute $\Ci_\fid$ in that case. In this work, we opt to retain the fiducial model, since it has several useful properties; most notably, the fact that the difference estimator (in the form of \eqref{eq: general-qe-est}) is able to cancel the leading-order effects of \resub{pixellization}, non-Poissonian shot-noise and unsampled bins.\footnote{To see this, consider the scenario in which our model for $\N$ is wrong by an additive factor. If using the form of \eqref{eq: qe-no-fid}, this will lead to a bias in $\hat{p}_\alpha$, but not from estimator \eqref{eq: general-qe-est} if it is contained within both $\C_D$ and $\C_\fid$, \textit{i.e.} if it is an effect present in the simulations. Essentially, we will not be biased by modeling inaccuracies in the fiducial covariance (as long as the derivatives $\C_{,\alpha}$ are correct, though these may lead to a slight loss of optimality.}

\section{Estimating Subspace Coefficients}\label{sec: subspace-qe}

The above sections have presented an extensive discussion concerning the application of quadratic estimators to measuring the unwindowed galaxy power spectrum from spectroscopic surveys. In general, this spectrum is high-dimensional (particularly in redshift-space), requiring significant computational power to compute the statistic and its sample covariance. For this reason, it is often useful to \textit{compress} the statistic, and a number of methods are available by which to perform this \citep[e.g.,][]{1997ApJ...480...22T,2000MNRAS.317..965H,2000ApJ...544..597S,2018MNRAS.476L..60A,2021PhRvD.103d3508P}. Here, we focus on the subspace compression introduced in Ref.\,\citep{2021PhRvD.103d3508P}, (which gives the optimal linear decomposition for a specific cosmological analysis) but note that the approach is applicable to any compression scheme in which the power spectrum can be written as a linear combination of basis functions.

\subsection{The Compressed Subspace}\label{subsec: sv-def}
We begin by recapitulating the approach of Ref.\,\citep{2021PhRvD.103d3508P}, in somewhat modified notation. The former work first computed the window-convolved power spectra via standard estimators, then projected these measurements into a subspace of reduced dimension, utilizing basis vectors computed from an SVD in the space of allowed theory models. Here, we work only with unbinned and unwindowed spectra, and compute the subspace coefficients directly, \textit{i.e.} without first computing the band-powers. This leads to a significant reduction in computational time.

The continuous-space decomposition is defined by first rotating an arbitrary spectrum $P(\vk)$ into a noise-weighted function $X(\vk)$, then decomposing into a set of $N_\mathrm{SV}$ basis functions $\{V^\alpha(\vk)\}$ with coefficients $\{c_\alpha\}$;
\beq\label{eq: subspace-def}
    X(\vk) &=& \int_{\vk'}C_\fid^{-1/2}(\vk,\vk')\left[P(\vk')-\bar{P}(\vk')\right]\\\nonumber
    X(\vk) &\approx& \sum_{\alpha=1}^{N_\mathrm{SV}}c_\alpha V^\alpha(\vk).
\eeq
In \eqref{eq: subspace-def}, $C_\mathrm{fid}$ is a fiducial covariance matrix of the power spectrum,\footnote{Note that we use $\C$ to indicate covariances of the pixellized data, and $C$ to refer to those of the power spectrum.} the square root indicates a Cholesky decomposition, and $\bar{P}(\vk)$ \resub{is a mean spectrum, ensuring that the average over all samples of $X(\vk)$ has mean zero.} This becomes exact in the limit of $N_\mathrm{SV}\rightarrow\infty$. In Ref.\,\citep{2021PhRvD.103d3508P}, the basis vectors $\{V^\alpha\}$ are chosen by drawing a `template bank' of noiseless power spectra from the theory model with some priors on cosmological and nuisance parameters, then performing an SVD on this set to define the basis vectors, with the mean of all templates additionally setting $\bar{P}$.\footnote{\resub{We note that choosing the basis vectors requires significantly more thought for the bispectrum, since an analogous SVD will not generate separable templates, which are required for fast computation.}} This template bank can also be used to define $N_\mathrm{SV}$, by enforcing some limit on the $\chi^2$ error incurred by the decomposition, averaged across the prior domain. In practice, the basis vectors are computed in a finite number of bins (and for the first few Legendre multipoles); these are easily converted to continuous form by way of interpolation. To define the decomposition, we must also set the fiducial power-spectrum covariance $C_\fid$; for an efficient and accurate decomposition, this should be a smooth model and similar to the true covariance $C_\mathrm{true}(\vk,\vk') = \operatorname{cov}\left(P(\vk),P(\vk')\right)$; in practice a Gaussian model is usually a good approximation \citep{2021PhRvD.103d3508P}.

Inverting \eqref{eq: subspace-def} allows the (unwindowed) power spectrum to be written as a linear sum of shapes;
\beq\label{eq: Pk-decomp}
    P(\vk) &\approx& \sum_{\alpha=1}^{N_\mathrm{SV}}c_\alpha \int_{\vk'}C_\mathrm{fid}^{1/2}(\vk,\vk')V^\alpha(\vk') + \bar{P}(\vk)\\\nonumber
    &\equiv& \sum_{\alpha=1}^{N_\mathrm{SV}} c_\alpha \omega^\alpha(\vk) + \bar{P}(\vk),
\eeq
defining the weighting functions $\omega^\alpha(\vk)$ in terms of the subspace basis vectors, $V^\alpha$, and the fiducial power-spectrum covariance, $C_\fid$. The basis vectors are orthonormal, satisfying $\int_{\vk}V^\alpha(\vk)V^\beta(\vk) = \delta^K_{\alpha\beta}$; this leads to the direct estimator for $c_\alpha$;
\beq\label{eq: c-alpha-conv}
    c_\alpha &=& \int_{\vk}\int_{\vk'}\left[P(\vk)-\bar{P}(\vk)\right]C_\mathrm{fid}^{-1/2}(\vk,\vk')V^\alpha(\vk')\\\nonumber
    &=& \int_{\vk}\int_{\vk'}\left[P(\vk)-\bar{P}(\vk)\right]C_\mathrm{fid}^{-1}(\vk,\vk')\omega^\alpha(\vk'),
\eeq
which is a convolution of $P(\vk)$ with the shape functions $\omega^\alpha$. This (in discrete form and applied to the windowed power spectrum multipoles), is the manner used to estimate $c_\alpha$ from theory and data in the former work. \resub{Since we only ever consider the difference between measured and theoretical $c_\alpha$ coefficients, we note that $\bar{P}(\vk)$ cancels in practical contexts.}

\subsection{Quadratic Estimators for $\{c_\alpha\}$}\label{subsec: sv-qe}

In this work, we estimate $\vec c \equiv \{c_\alpha\}$ directly using the quadratic estimators, rather than via \eqref{eq: c-alpha-conv}. To do so, we first express \eqref{eq: Pk-decomp} in terms of multipoles, defining the new weightings $\omega_\ell^\alpha(k)$;
\beq
    P(\vk;\vr') \approx \sum_\ell \left[\sum_{\alpha=1}^{N_\mathrm{SV}} c_\alpha \omega_\ell^\alpha(k) + \bar{P}_\ell(k)\right]L_\ell(\hat{\vk}\cdot\hat{\vr}')
\eeq
allowing for dependence on the line-of-sight $\hat{\vr}'$, as in the Yamamoto estimator. The $\omega_\ell^\alpha(k)$ functions are, in principle, arbitrary, though we will use them as the interpolated (discrete) SVD basis vectors, \textit{i.e.} $\omega_\ell^\alpha(k_i) = \omega_{\ell,i}^\alpha$ where $i$ indexes the $k$-bin used to form the template bank. By full analogy with Sec.\,\ref{sec: qe-theory}, the quadratic estimator for $\vec{c}$ is given by \eqref{eq: quadratic-estimator-def} or \eqref{eq: general-qe-est}, but replacing the $\C_{,\alpha}\equiv\partial \C / \partial P_\alpha$ with $\partial \C/\partial c_\alpha$. This is a straightforward redefinition of \eqref{eq: C-alpha-def};
\beq\label{eq: C-alpha-sv}
    \C_{,\alpha}(\vr,\vr')\rightarrow\frac{\partial \C}{\partial c_\alpha}(\vr,\vr') = n(\vr)n(\vr')\int_{\vk}e^{i\vk\cdot(\vr-\vr')}\sum_\ell\omega_\ell^\alpha(k)L_\ell(\hat{\vk}\cdot\hat{\vr}') 
\eeq
inserting $\omega^\alpha_\ell$ in place of the filter functions $\Theta^a(\vk)$ (which were simply Fourier space shells). Note that \eqref{eq: C-alpha-sv} retains the sum over $\ell$, since the subspace basis vectors contain contributions from all multipoles used to compute the create the template bank. 
Additionally, the wavenumber support of $\omega_\ell^\alpha(k)$ can be set by truncating the template bank spectra at some $k_\mathrm{min}$ and $k_\mathrm{max}$. The application of \eqref{eq: C-alpha-sv} on pixellized fields is analogous to \eqref{eq: applyC-alpha} and we note that there is no change to the pixel weighting $\mathsf{H}_\fkp$ or $\C_\fid$ (with the latter already depending on the full power spectrum in the fiducial cosmology).

To use the quadratic estimator, we additionally require the fiducial coefficients $c_\alpha^\mathrm{fid}$. From \eqref{eq: c-alpha-conv}, these are defined by
\beq\label{eq: fid-coeff}
    c_\alpha^\mathrm{fid} &=& \int\frac{d\vr'}{V}\int_{\vk}\tilde{\omega}^\alpha(\vk;\vr')\left[P(\vk;\vr')-\bar{P}(\vk;\vr')\right]\\\nonumber
\eeq
averaging over the line-of-sight. This uses the inverse weights, defined as in \eqref{eq: c-alpha-conv}; 
\beq
    \tilde{\omega}^\alpha(\vk;\vr') &=& \int_{\vk'}C^{-1}_\fid(\vk,\vk')\omega^\alpha(\vk';\vr')\\\nonumber
     &=& \sum_\ell (2\ell+1)\tilde{\omega}_\ell^\alpha(k)L_\ell(\hat{\vk}\cdot\hat{\vr}'),
\eeq
with a multipole decomposition given in the second line. As in Sec.\,\ref{subsec: implement-fid}, there are two possibilities for computing $c_\alpha$, either allowing for or removing discreteness effects. In the first case, we estimate the $\tilde{\omega}_\ell^\alpha$ functions from the discretely sampled SVD vectors via $\tilde{\omega}^\alpha_\ell(k_i) = (2\pi^2)/(k_i^2\delta k_i)\times \left[C_\mathrm{fid}^{-1}\omega^\alpha\right]_{\ell,i}$, with the prefactor necessary to ensure correct normalization. This gives the fiducial coefficients analogous to \eqref{eq: p-fid-discrete};
\beq
    c_\alpha^{\fid,\mathrm{discrete}} = 4\pi\left[\int_{\vk}\sum_{\ell,L} \frac{4\pi}{2L+1}\left[P_L(k)-\bar{P}_L(k)\right]\tilde{\omega}^\alpha_\ell(k) Y^*_{\ell m}(\hat{\vk})Y^{}_{LM}(\hat{\vk})\int \frac{d\vec r'}{V}Y_{\ell m}(\hat{\vec r}')Y^*_{LM}(\hat{\vec r}')\right]
\eeq
where the integral is evaluated as a sum over all points on the real and Fourier-space grids. 

The alternative approach, which will be used here, is to perform the angular integrals in \eqref{eq: fid-coeff} analytically, leading to
\beq
    c_\alpha^\fid \rightarrow \int \frac{k^2dk}{2\pi^2}\tilde{\omega}^\alpha(k)\left[P_\ell(k)-\bar{P}_\ell(k)\right].
\eeq
Given the SVD basis vectors $V^\alpha$ in discrete form, this becomes a matrix multiplication of the fiducial spectrum $P_\ell(k)$ (evaluated at the same wavenumbers used to define the SVD) and $\tilde\omega_i = \sum_j C_{\fid,ij}^{-1/2}V_{j}^\alpha$. This approach eliminates leading-order effects of gridding artefacts and the finite-$k$ sampling used to define the basis vectors.

Following the above steps, we obtain a practical estimator for computing the unwindowed subspace coefficients directly from the pixellized data, without first computing the band-powers. In this context, the subspace decomposition has dual utility; first, the associated dimensionality reduction (\textit{i.e.} $N_\mathrm{SV}\ll N_\mathrm{bins}$) reduces the number of mocks required to compute sample covariance matrices (as in \cite{2021PhRvD.103d3508P}), secondly, it significantly expedites the estimator itself (both for the ML and FKP pixel weights). This occurs since the number of FFTs required to compute the statistic scales as $N_\mathrm{mc}N_\mathrm{bins}(N_\mathrm{it}+N_\mathrm{bins})$ (Sec.\,\ref{subsec: implement-bias-fish}), with $N_\mathrm{bins}\rightarrow N_\mathrm{SV}$ in the compressed formalism.

This approach carries an important subtlety. As previously noted, the quadratic estimator is fully unbiased only if we simultaneously estimate all possible band-powers $\vp$, or equivalently, coefficients $\vec c$. Using only a subset incurs a small error unless the unsampled coefficients are (a) independent from those sampled or (b) the same in the data and the fiducial model.\footnote{Schematically, this can be seen from writing \eqref{eq: bias2} as $\avEst{\hat{p}_\alpha}-p^\fid_\alpha =  \sum_{\beta,\gamma}F_{\alpha\beta}^{-1}\mathcal{F}^{}_{\beta\gamma}(p_\gamma^\mathrm{true}-p^\fid_\gamma)$, where $F$ is a submatrix of the full (infinite-dimensional) Fisher matrix $\mathcal{F}$ and $\beta$ runs only over the estimated coefficients. If the unsampled bins are uncorrelated then $\sum_{\beta}F_{\alpha\beta}^{-1}\mathcal{F}^{}_{\beta\gamma} = \delta^K_{\alpha\beta}$, giving $\hat{p}_\alpha = p^\mathrm{true}_\alpha$, and similarly if $p_\gamma^\mathrm{true}=p^\fid_\gamma$ for those coefficients.} When estimating band-powers this is of little concern, since most unmeasured bins contribute little power in practice, given the lack of a window function; for the coefficients however, restricting to small $N_\mathrm{SV}$ can lead to a bias and increased variance in the estimator if there remains significant information outside those coefficients. (Recall that the subspace basis vectors are chosen to maximize information content in the final cosmological analysis, not that the maximum signal-to-noise in the spectrum). In practice, we find this to be the case when using the FKP quadratic estimator for the subspace coefficients (which is less efficient than the ML approach at removing window-function effects), and ameliorate it by measuring twice the number of coefficients, with the latter half discarded when the coefficients are later analyzed via MCMC. One can straightforwardly assess whether this effect is present by using only a subset of the coefficients in the estimated $\hat{q}_\alpha$ and $F_{\alpha\beta}$ quantities \eqref{eq: general-qe-est} and checking for variation in $\hat{p}_\alpha$.

\section{Demonstration on BOSS DR12}\label{sec: boss-analysis}

As a practical demonstration of the quadratic estimators discussed herein, we estimate the power spectrum multipoles and compressed subspace coefficients of a real cosmological data-set. For this we use the twelfth data-release (DR12) \citep{2017MNRAS.470.2617A} of the Baryon Oscillation Spectroscopic Survey (BOSS), part of SDSS-III \citep{2011AJ....142...72E,2016arXiv161100036D}, specializing to the patch with the largest number density (and volume); the north Galactic cap in the redshift range $0.2<z<0.5$, with total volume $1.46h^{-3}\mathrm{Gpc}^3$ and mean redshift $z = 0.38$. 

A set of MultiDark-Patchy (hereafter `Patchy') simulations\footnote{Publicly available at \href{https://data.sdss.org/sas/dr12/boss/lss/}{data.sdss.org/sas/dr12/boss/lss}.} \cite{2016MNRAS.460.1173R,2016MNRAS.456.4156K} are used to define the fiducial cosmology, which, by design, have power spectra close to those of BOSS and share its selection function. For each simulation, we compute the windowed power spectrum multipoles using the estimator of Ref.\,\citep{2017JCAP...07..002H} within \texttt{nbodykit} \citep{2018AJ....156..160H} (using the same gridding parameters as for the public BOSS data release), then fit the mean of 1000 simulations to the one-loop EFT model \citep{2020JCAP...05..042I}, using the \texttt{CLASS-PT} implementation \cite{2020PhRvD.102f3533C}. For the fit, we use $k\in[0.01,0.25]\hMpc$ and $\ell\in\{0,2\}$, fixing the cosmological parameters to their true values ($\{\Omega_m = 0.307115, \Omega_b = 0.048206, \sigma_8 = 0.8288, n_s = 0.9611, h = 0.6777, A_s = 2.1467\times 10^{-9}\}$) and varying only the nuisance parameters, encapsulating linear and non-linear \resub{galaxy} bias, stochasticity and the EFT counterterms. Finally, the fiducial power spectrum is set equal to the best-fit theory model,  \resub{computed using the public window-function multipoles and} subtracting Poisson shot-noise; \resub{the latter} is necessary to ensure that we do not double-count stochasticity in both the signal and noise covariance of the data. We require also a smooth model of the power spectrum for $k>k_\mathrm{max}$; this is set to linear theory, rescaled to avoid discontinuities at $k_\mathrm{max}$. The exact choice is of only minor importance, since these modes are not used in the later analysis.

\subsection{Estimating Unwindowed Spectra}
Using the above, we may apply the quadratic estimators discussed in Sec.\,\ref{sec: qe-theory}\,\&\,\ref{sec: qe-implementation} to measure the band-powers of both data and simulations via \eqref{eq: general-qe-est}. We will assume the binning $k_\mathrm{min}=0\hMpc$, $k_\mathrm{max} = 0.3\hMpc$, $\Delta k = 0.005\hMpc$ and $\ell \in \{0,2\}$, somewhat broader than the $k$-range used for the subsequent MCMC analysis to avoid the aforementioned bias from correlated unmeasured bins. For both ML and FKP pixel weights, we require a set of galaxy positions and random particles; for the latter, we use the publicly available catalogs\footnote{Also at \href{https://data.sdss.org/sas/dr12/boss/lss/}{data.sdss.org/sas/dr12/boss/lss/}.} generated for the Patchy and BOSS samples which have 50$\times$ the average density of the galaxy samples.\footnote{See Appendix \ref{appen: pix-options} for the effects of reducing this to $10\times$.} We include the full particle weights of BOSS (including completeness, systematic and fiber collision weights, but not FKP particle weights),\footnote{When using the FKP pixel weights (\textit{i.e.} $\mathsf{H} = \mathsf{H}_\fkp$), we additionally divide the FKP Fisher matrix by a factor $\int d\vr\,n^2(\vr) \times \alpha /\sum_i \bar{n}_i$, where $\bar{n}_i$ is the number density associated with the random catalog and $i$ runs over all random particles. This corrects for differences in normalization of the gridded and windowed-FKP estimators when only a finite number of randoms are used.} and translate data and randoms onto Cartesian grids using the fiducial cosmology $\{\Omega_m = 0.31, h = 0.676\}$ (as in BOSS analyses). Gridding is achieved according to triangular-shaped-cloud interpolation utilizing the same cuboidal grids as for the public $P(k)$ spectra, but with the number of pixels in each dimension optionally reduced by a factor, $f_\mathrm{pix}$, of two or three.\footnote{The dependence of the spectra on $f_\mathrm{pix}$ is discussed in Appendix \ref{appen: pix-options}.} These have associated Nyquist frequencies $k_\mathrm{Nyq} \approx (0.6/f_\mathrm{pix})\hMpc$. We precompute the fiducial power spectrum multipoles $P_\ell(\vk)$ and the spherical harmonics $Y_{\ell m}(\hat{\vk})$ and $Y_{\ell m}(\hat{\vr})$ on the grid to facilitate later computations. 

With these pixellized fields, we compute the quadratic estimator terms $\vd^T\Hi \C_{,\alpha}\Hi\vd$ and, for the ML case, $\vd^T\Hi \C_{,\alpha}\Hi\C_{,\beta}\Hi\vd$ as detailed in Sec.\,\ref{sec: qe-implementation}. For the FKP case, the Fisher matrix is instead computed using subsets of the random catalog, as in Sec.\,\ref{subsec: implement-bias-fish}. This operation is performed both for the data and for 200 Patchy mocks (and using 50 random partitions to define the FKP Fisher matrix); 50 of these are used to define the bias and Fisher matrix via Monte Carlo averaging,\footnote{As in Sec.\,\ref{subsec: implement-bias-fish}, using 50 simulations leads to a $\approx 1\%$ increase in the estimator variance relative to that of infinite mocks.} allowing the estimated BOSS band-powers $\hat{p}_\alpha$ to be computed via \eqref{eq: general-qe-est} or \eqref{eq: quadratic-estimator-def}; the remainder are used to form a covariance matrix of the statistic. Whilst this gives a good indication of the statistic correlation structure, it is insufficient for performing parameter inference, as one requires $N_\mathrm{mocks} \gg N_\mathrm{bins}$ to obtain a robust inverse covariance matrix estimate. For this reason, we run MCMC only on the compressed subspace vectors (discussed below). Using the ML pixel weights, each simulation requires $\sim 0.5$ ($1.5$) hours to analyze on a four-core machine for $k_\mathrm{Nyq} = 0.2$ ($0.3$) $\hMpc$; for the FKP case, we require $\sim 1$ ($0.5$) minutes to compute the bias term in each simulation, though with an additional $\sim 0.5$ ($1.5$) hours for each Fisher matrix random catalog partition. As previously discussed, this is similar to the ML case since the number of bins is large.


\begin{figure}
    \centering
    \includegraphics[width=0.6\textwidth]{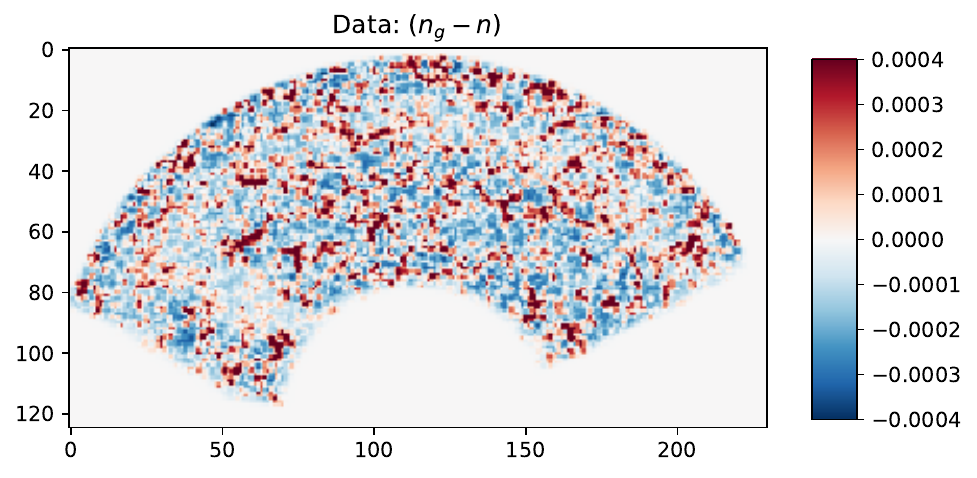}\\
    \includegraphics[width=0.45\textwidth]{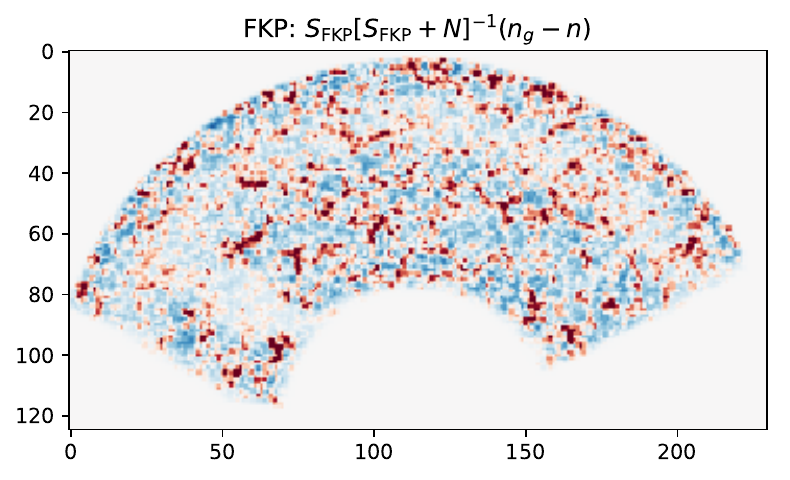}
    \includegraphics[width=0.45\textwidth]{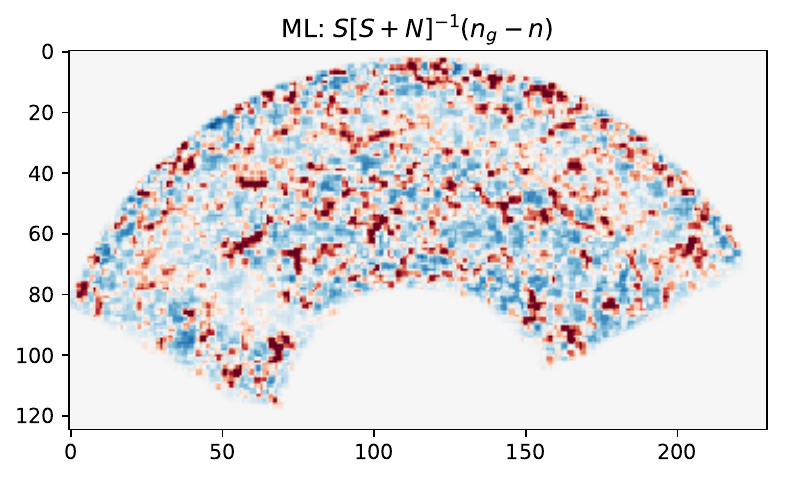}
    \caption{Visualizations of the overdensity field for a single MultiDark-Patchy mock, both unweighted and applying Wiener filters using the FKP \eqref{eq: H-fkp-def} and maximum-likelihood (ML) \eqref{eq: quadratic-estimator-def} pixel covariances discussed in this work. Each image depicts a slice of radial width $\Delta z \approx 100\Mpch$, with horizontal and vertical axes labeling pixels (in the transverse direction) with grid-size $\approx 11\Mpch$ per pixel. We apply the Wiener filters $\Sig[\Sig+\N]^{-1}$ for signal and noise covariances $\Sig$ and $\N$ rather than the $[\Sig+\N]^{-1}$ weights used for power spectrum estimation to facilitate easier interpretation. The ML covariance is inverted using conjugate gradient descent methods described in Sec.\,\ref{subsec: implement-C-Ca}.}
    \label{fig: wiener-plot}
\end{figure}

Before discussing the estimated power spectrum multipoles, it is instructive to consider the action of the pixel weights in the FKP and optimal formalism. Fig.\,\ref{fig: wiener-plot} shows a slice of the 3D overdensity field $d(\vr)\equiv n_g(\vr)-n(\vr)$, alongside its Wiener-filtered form $\Sig[\Sig+\N]^{-1}\vd$, using the FKP and ML signal covariances $\Sig$. Even though the raw data does not include any selection-based weights, the differences between the unfiltered and filtered maps is relatively small, though we note that the former contains more small-scale power. The differences between FKP and ML weights are again slight, though the ML scheme appears to upweight long-wavelength modes. Whilst this discussion is somewhat rudimentary, it suggests that the differences between the ML and FKP spectra may be small, except on the largest scales.

\begin{figure}
    \centering
    \includegraphics[width=0.8\textwidth]{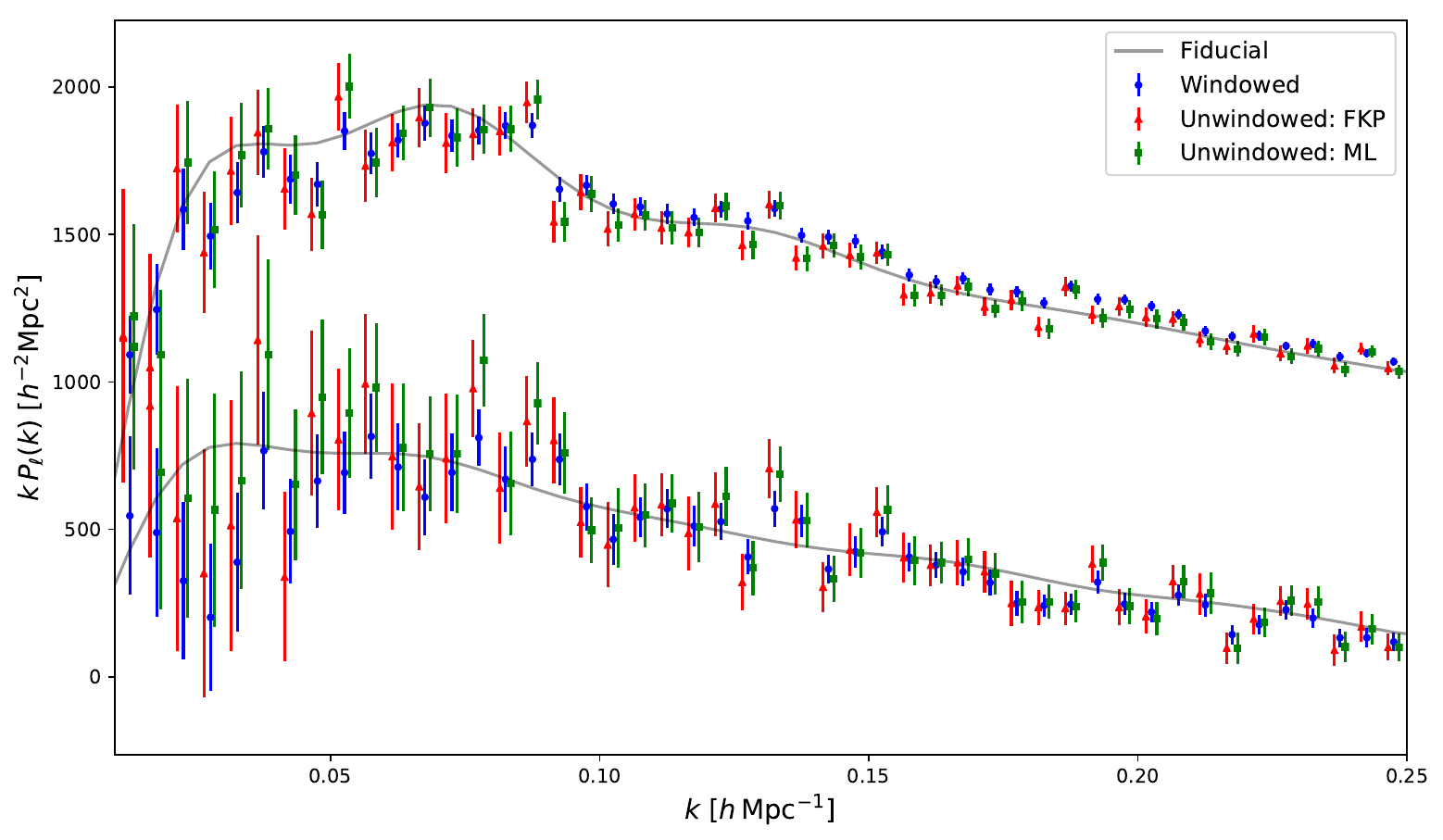}
    \caption{Galaxy power spectra for the largest-volume chunk of the BOSS DR12 data-set, \resub{computed} both using the conventional algorithm available in \texttt{nbodykit} (blue) and the quadratic estimators discussed in this work using FKP (red) and maximum-likelihood (ML, green) pixel weightings. Only the first set of spectra are convolved with the survey window function, though neither include Poissonian shot-noise. Data is obtained by applying the formalism of Secs.\,\ref{sec: qe-theory}\,\&\,\ref{sec: qe-implementation} to the observed BOSS data in the NGC patch with $0.2<z<0.5$, using 150 Patchy simulations to compute the variances. The quadratic estimators grid the data with a Nyquist frequency $k_\mathrm{Nyq} = 0.3\hMpc$ (which is sufficient for this method), and use a fiducial power spectrum shown by the grey lines. The upper (lower) data-points show the redshift-space monopole (quadrupole) and we add small lateral displacements for clarity. Whilst the windowed estimates have reduced variance at high-$k$, individual bins are highly correlated, as shown in Fig.\,\ref{fig: pk-cov-plot}.}
    \label{fig: pk-full-plot}
\end{figure}

Fig.\,\ref{fig: pk-full-plot} shows the band-power estimates from the two quadratic estimators alongside the standard (windowed) power spectrum multipoles computed using \texttt{nbodykit}. The quadratic estimators are computed using a grid-size corresponding to $k_\mathrm{Nyq} = 0.3\hMpc$; whilst this may seem insufficiently fine (indeed, we set $k_\mathrm{Nyq} = 0.6\hMpc$ for the \texttt{nbodykit} estimates to avoid aliasing), using broad cells makes little difference to the band-power estimates due to the form of \eqref{eq: general-qe-est} as a difference estimator. This is elaborated upon in Appendix \ref{appen: pix-options}, displaying different choices of $k_\mathrm{Nyq}$. 

Firstly, we note that, whilst the spectra from the quadratic estimators are not directly compatible with those from \texttt{nbodykit} due to the removal of window function effects, they are heuristically similar, and notably, both show the same departures from the fiducial power spectrum model. Especially at high-$k$, it is clear that the windowed estimates have smaller variances than their unwindowed equivalents; this is not an indication that the quadratic estimators are incomplete, but rather due to the presence of significant correlations between bins induced by the window function. Indeed, the total signal-to-noise of the two approaches is very similar. Comparing now the quadratic estimators with FKP and ML pixel weights, we find very similar results, both for the means and variances of each band-power, indicating that the FKP form is close to the optimal Gaussian solution for this choice of bins. Further discussion of this may be found in Sec.\,\ref{subsec: ml-fkp-comparision}.

\begin{figure}
    \centering
    \includegraphics[width=0.8\textwidth]{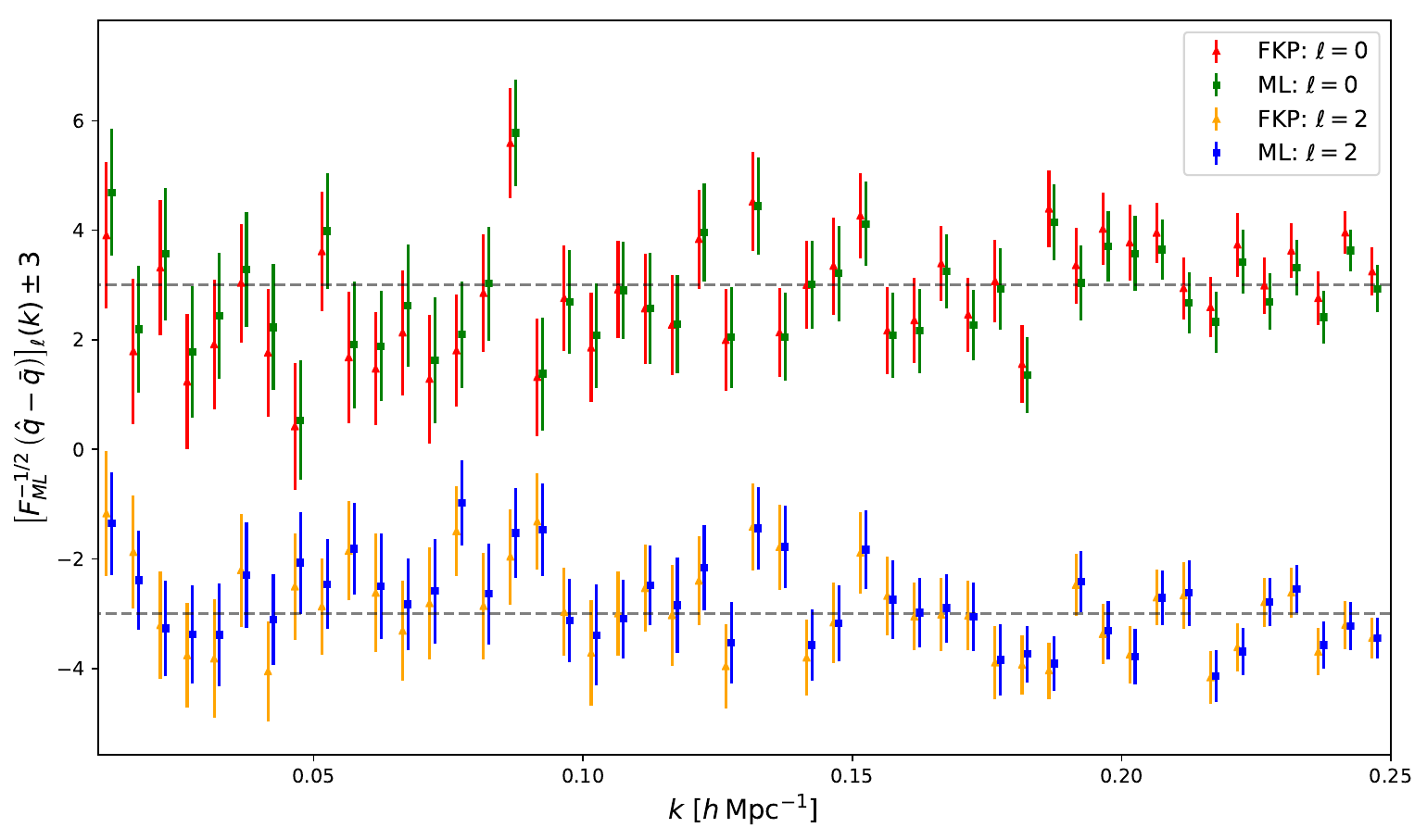}
    \caption{Unwindowed BOSS power spectrum estimates from the quadratic estimators, rescaled to the convention of Ref.\,\citep{2004ApJ...606..702T}. We plot the quantity $\hat{q}_\alpha-\bar{q}_\alpha$ (where $\hat{q}_\alpha$ is the part of the estimator containing the data and $\bar{q}$ is its equivalent averaged over mocks), normalized by the Cholesky decomposition of the inverse ML Fisher matrix; this combination has close to unit covariance matrix for the ML sample. The top (bottom) lines show result for the monopole (quadrupole) and we note consistent estimates and similar error-bars in both cases, with only a slight improvement seen from the FKP estimators at low-$k$.}
    \label{fig: pk-rescaled-plot}
\end{figure}

Given that individual band-powers will have a non-trivial correlation structure (even in the ML case), a more appropriate quantity to plot is the vector $\sum_{\beta}F_{\alpha\beta}^{-1/2}\left[\hat{q}_\beta-\bar{q}_\beta\right]$, which includes all contributions from the data, and, for the ML estimator, has unit covariance in the Gaussian limit \citep{2004ApJ...606..702T}. This is plotted in Fig.\,\ref{fig: pk-rescaled-plot} and allows for better comparison of the two approaches. Notably, the variance of the ML estimates is close to unity for all the bins, implying that the estimator is near-optimal. Once again we find little difference between the FKP and ML approaches, except for a possible slight reduction in the errorbars from the ML estimators on the largest scales (which would be useful, for example, for surveys focusing on primordial non-Gaussianity). A more meaningful comparison of the spectra is via their resulting cosmological parameter constraints; these will be discussed in Sec.\,\ref{subsec: param-inference}.

\begin{figure}
    \centering
    \includegraphics[width=0.8\textwidth]{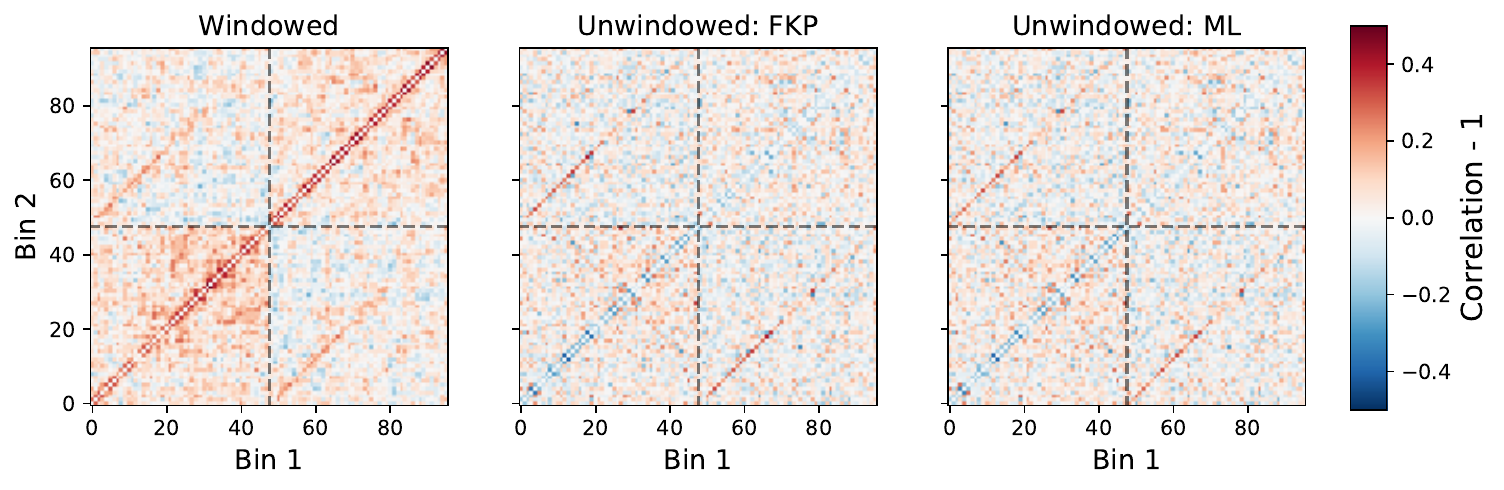}
    \caption{Correlation matrices for the power spectra plotted in Fig.\,\ref{fig: pk-full-plot}, defined as $\operatorname{corr}_{ab} = \mathrm{cov}_{ab}/\sqrt{\mathrm{cov}_{aa}\mathrm{cov}_{bb}}$ for binned covariance $\mathrm{cov}_{xy}$. These are generated from 150 Patchy mocks, and bins $0-47$ ($48-95$) refer to the monopole (quadrupole) with increasing $k$ in the range $[0.01,0.25]\hMpc$. For clarity, we subtract the identity matrix to remove the leading diagonal. We note considerably correlation between neighbouring bins and at high-$k$ for the windowed spectra which is significantly reduced using the quadratic estimators discussed in this work, though there is a slight anti-correlation of neighbouring bins.}
    \label{fig: pk-cov-plot}
\end{figure}

Finally, Fig.\,\ref{fig: pk-cov-plot} considers the correlation matrix of the above power spectra, computed from the 150 Patchy mock band-power estimates. The windowed spectra exhibit clear positive correlations for off-diagonal elements, particularly those in neighbouring bins, but also pairs at larger separations on small scales. In contrast, the unwindowed quadratic estimator covariances are almost diagonal (and very similar for FKP and ML pixel weights), with just a slight hint of positive correlations at high-$k$, and a slight anticorrelation between adjacent bins. This is as expected; the window function smears the true power spectrum covariance into nearby bins, which also accounts for the reduced variance seen at high-$k$ in Fig.\,\ref{fig: pk-full-plot}.

\subsection{Direct Computation of Subspace Coefficients}\label{subsec: subspace-results}
Fast and accurate analyses of galaxy power spectra are facilitated by projecting the spectra into information-maximizing subspaces, as in Ref.\,\cite{2021PhRvD.103d3508P}. This is is possible via the algorithms discussed in Sec.\,\ref{sec: subspace-qe} and allows for robust parameter inferences using only $\mathcal{O}(100)$ mocks. This is far more computationally feasible, given the increase in computing power needed for estimating spectra via configuration-space quadratic estimators.

To compute the compressed spectra, we first require a set of basis vectors, generate from a template bank of theory models given prior ranges on cosmological and nuisance parameters. Unlike Ref.\,\cite{2021PhRvD.103d3508P}, we do not need to include window functions, and can use dense $k$-space sampling to ensure information is not lost in binning. We sample $10^4$ template bank parameters from the following distributions (varying just $h, \omega_\mathrm{cdm}$, $A_s$ and nuisance parameters, for simplicity);
\beq\label{eq: template-bank-params}
    h &\sim& \mathcal{U}(0.6,0.74),\quad \omega_\mathrm{cdm} \sim \mathcal{U}(0.08,0.16),\quad (A_s/A_{s,\mathrm{Planck}})^{1/2} \sim \mathcal{U}(0.6,1.4),\\\nonumber
    b_1 &\sim& \mathcal{U}(1.7,2.3), \quad b_2 \sim \mathcal{N}(0,1^2)\quad b_{\mathcal{G}_2} \sim \mathcal{N}(0,1^2),\\\nonumber
    c_{s,0}&\sim& \mathcal{N}(0,30^2),\quad c_{s,2}\sim \mathcal{N}(0,30^2),\quad P_\mathrm{shot}\sim \mathcal{N}(0,5000^3),\\\nonumber
    b_4&\sim& \mathcal{N}(500,500^3),
\eeq
where $\mathcal{U}$ and $\mathcal{N}$ represent uniform and normal distributions. Spectra are evaluated for $k\in[0.01,0.25]\hMpc$ with $\Delta k = 0.0005\hMpc$ and $\ell\in\{0,2\}$, choosing the $k$-range to limit the impact of observational systematics and model inaccuracies. Furthermore, we set the fiducial covariance equal to a simple Gaussian model with true survey volume, and include a theoretical error covariance \cite{2016arXiv160200674B} of the form
\beq\label{eq: cov-te}
    C_{\mathrm{TE}, \ell\ell'}(k,k') = E_\ell(k)E_{\ell'}(k')e^{-(k-k')^2/2\Delta k^2}
\eeq
where $E_\ell(k) = D^2(z)P^\mathrm{tree}_\ell(k)(k/0.45)^{3.3}$, similar to that used in Ref.\,\cite{2019JCAP...11..034C}. Here the coherence length $\Delta k = 0.1\hMpc$ ensures that we still retain BAO information from small scales, but are robust to broadband modeling uncertainties. Given the template bank, the subspace is formed via an SVD, and we use only the first 30 basis vectors.\footnote{\resub{As in Ref.\,\citep{2021PhRvD.103d3508P}, we choose $N_\mathrm{SV}$ by requiring that the basis projection incurs an error in $\chi^2$ (averaged across the prior domain) below $0.1$. This gives $N_\mathrm{SV}\geq 15$, which the former work showed to capture essentially all the cosmological information available.} Here, we estimate twice the number of coefficients to minimize the estimator bias from unsampled modes (cf.\,Sec.\,\ref{subsec: sv-qe}. Spectra obtained using $N_\mathrm{SV} = 15$ are shown in Appendix \ref{appen: pix-options}.} Following computation of the basis vectors (which we emphasize is fast, requiring only $10^4$ model evaluations, each of which requires one \texttt{CLASS} evaluation, and a negligibly fast evaluation of the \texttt{CLASS-PT} module), we compute the subspace coefficients directly, as in Sec.\,\ref{sec: subspace-qe}, utilizing $N_\mathrm{mc}=50$ simulations for the Monte Carlo averages and $N_\mathrm{sim}=150$ for the covariances. \resub{These numbers are more than sufficient, given that the error on the Fisher matrix scales as $\sqrt{1+1/N_\mathrm{mc}}$ and the error inflation from using too few mocks scales roughly as $(N_\mathrm{SV}-N_\mathrm{param})/N_\mathrm{sim}$ for an analysis measuring $N_\mathrm{param}$ parameters. The compressed analysis is} significantly faster than that for the unwindowed spectra since fewer FFTs are required; it takes $\sim 7$, $20$ and $220$ minutes per simulation on a four-core machine using $k_\mathrm{Nyq} = 0.2, 0.3$ and $0.6$ $\hMpc$ respectively with ML pixel weights (and a comparatively similar time with FKP weights, but only to compute the Fisher matrix from the random particle subsamples).

\begin{figure}
    \centering
    \centering
    \subfloat{{\includegraphics[width=0.45\textwidth]{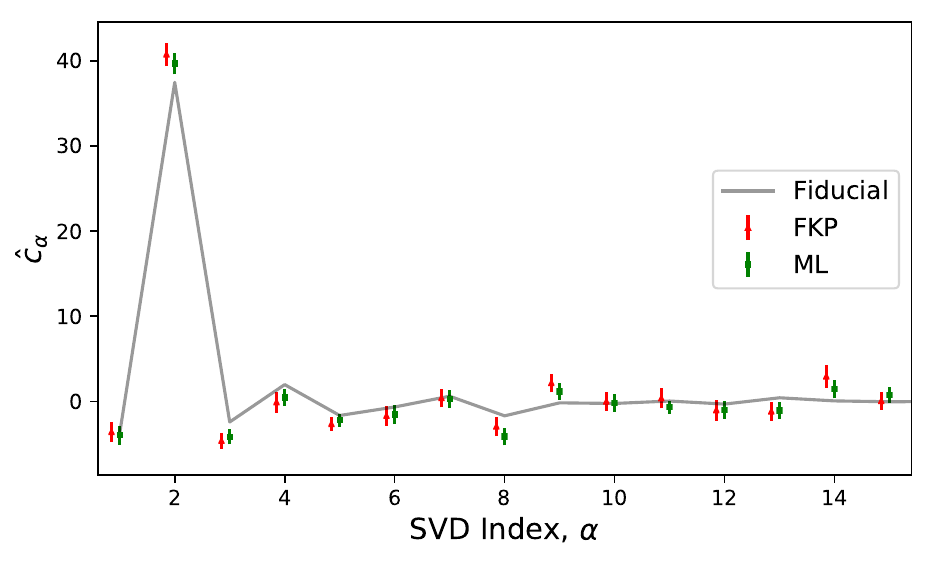} }}%
    \qquad
    \subfloat{{\includegraphics[width=0.45\textwidth]{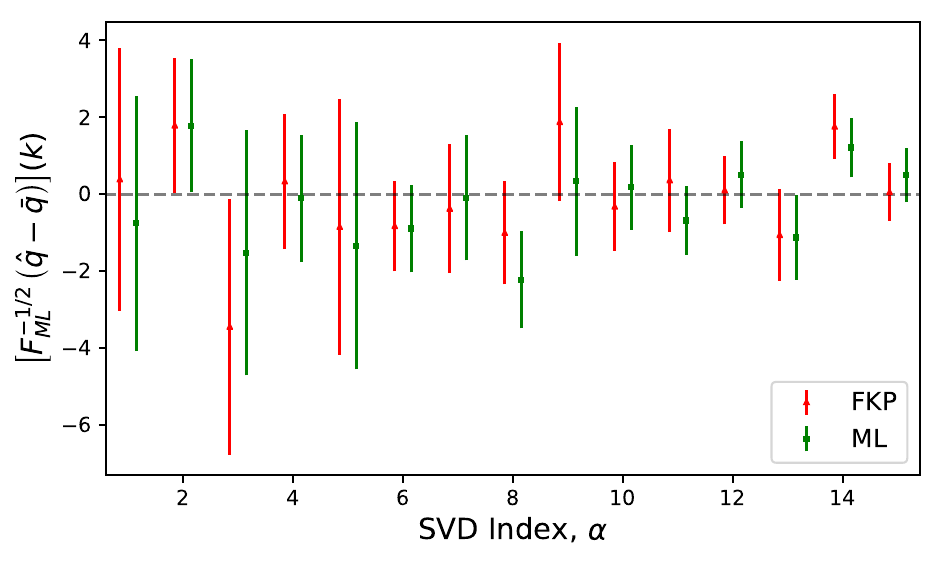} }}%
    \caption{Compressed coefficients of the BOSS DR12 power spectrum, obtained using the quadratic estimators of Sec.\,\ref{sec: subspace-qe} with FKP (green) and ML (red) pixel weights. The left panel shows the full coefficients alongside the fiducial model in gray (analogous to Fig.\,\ref{fig: pk-full-plot}), whilst the right employs the rescaling of Fig.\,\ref{fig: pk-rescaled-plot} to reduce correlations between bins. \resub{For the left panel, we note that only the \textit{difference} of fiducial and estimate $\hat{c}_\alpha$ is of interest; the offset cancels in any likelihood.} Points show the estimates obtained from the observed BOSS density field, whilst the errorbars give the covariances from 150 Patchy mocks. A total of 30 coefficients are estimated, but only the first 15 (shown here) are used for the later parameter inference. Each represents the amplitude of a basis vector in the subspace described in Sec.\,\ref{subsec: subspace-results}. We find similar results from the FKP and ML estimators, as in previous figures. }
    \label{fig: subspace-coeff}
\end{figure}

The first 15 subspace coefficients as shown in Figs.\,\ref{fig: subspace-coeff}, both plotted directly and rescaled in the manner of Fig.\,\ref{fig: pk-rescaled-plot}. The narrative is similar to that of the band-powers; the coefficients exhibit some variation around the fiducial coefficients calculated in the Patchy cosmology, but are is consistent between the two choices of weightings, with similar error bars. Notably, the amplitudes for the coefficients are generally close to zero with variances close to unity; this is expected from the SVD decomposition, as discussed in Ref.\,\citep{2021PhRvD.103d3508P}. The large amplitude of the second coefficient will relate to a significant difference between the mean of the basis vector template bank cosmology and that of the Patchy (and true) cosmology; in general, each basis vector depends on several different parameters (both cosmological and nuisance) and their priors, making their direct interpretation difficult. \resub{We re-emphasize that, since data and theory model are compressed in the same way, any offsets in $\hat{c}_\alpha$ do not enter the likelihood. By changing the strategy by which the template samples are drawn, we can modify the mean $\hat{c}_\alpha$; any such change does not bias the inference however (though the compression requires fewest basis vectors if the template samples are drawn from the parameter posteriors of the data).}

\subsection{Parameter Inference Without Windows}\label{subsec: param-inference}
We finally consider the performance of the quadratic estimators in their main application; cosmological parameter inference. For this purpose, we run a simple analysis varying the parameter set $\{h, \omega_\mathrm{cdm}, A_s\}$ alongside the nuisance parameters, using the likelihoods of \citep{2020JCAP...05..042I},\footnote{These are publicly available at \href{http://github/michalychforever/lss_montepython}{github.com/michalychforever/lss\_montepython}.} based on \texttt{CLASS-PT} \citep{2020PhRvD.102f3533C}. These include the analytic marginalization procedures of Ref.\,\cite{2021PhRvD.103d3508P} for parameters that enter the likelihood linearly (the residual shot-noise $P_\mathrm{shot}$ and the counterterms $c_0$, $c_2$, $b_4$), with only six parameters (the above plus $\{b_1, b_2, b_{\mathcal{G}_2}\}$ being directly sampled. We focus on the analysis of subspace coefficients with both ML and FKP pixel weights (utilizing a grid with $k_\mathrm{Nyq} = 0.3\hMpc$), noting that the conclusions will apply also to the full unwindowed band-powers.

Inference is performed via MCMC using the \texttt{montepython} sampler \citep{2019PDU....24..260B}, comparing the observed subspace coefficients $\hat{c}_\alpha$ to those from the finely-sampled one-loop power spectrum model with $k\in[0.01,0.25]\hMpc$, projected into the subspace via the aforementioned basis vectors. Since the basis vectors are set to zero outside this $k$-range, we do not have contributions from poorly modelled $k$-bins. For the covariance matrix, we use the sample covariance computed from 150 Patchy mocks supplemented with the theoretical error covariance $C_\mathrm{TE}$ of \eqref{eq: cov-te} and do not apply any Hartlap or Sellentin-Heavens noise-corrections \cite{2007A&A...464..399H,2017MNRAS.464.4658S} since $N_\mathrm{mocks}\gg N_\mathrm{bins}$. Broad Gaussian priors are applied to the nuisance parameters (\resub{barring} $b_1$) with widths given in \eqref{eq: template-bank-params} but centered on the values in the fiducial spectra. We run three MCMC chains: one for each of the FKP- and ML-weighted quadratic estimators discussed in this work, and a third using the full windowed power spectrum estimates, with a sample covariance drawn from 1000 Patchy mocks. For the latter, both $C_\mathrm{TE}$ and the theory model (at each step in the chain) must be convolved with the window function. All chains are run until convergence (assumed when the Gelman-Rubin diagnostic satisfies $|R-1|<0.05$), which requires $\sim 50$ CPU-hours per inference.

\begin{figure}
    \centering
    \includegraphics[width=0.8\textwidth]{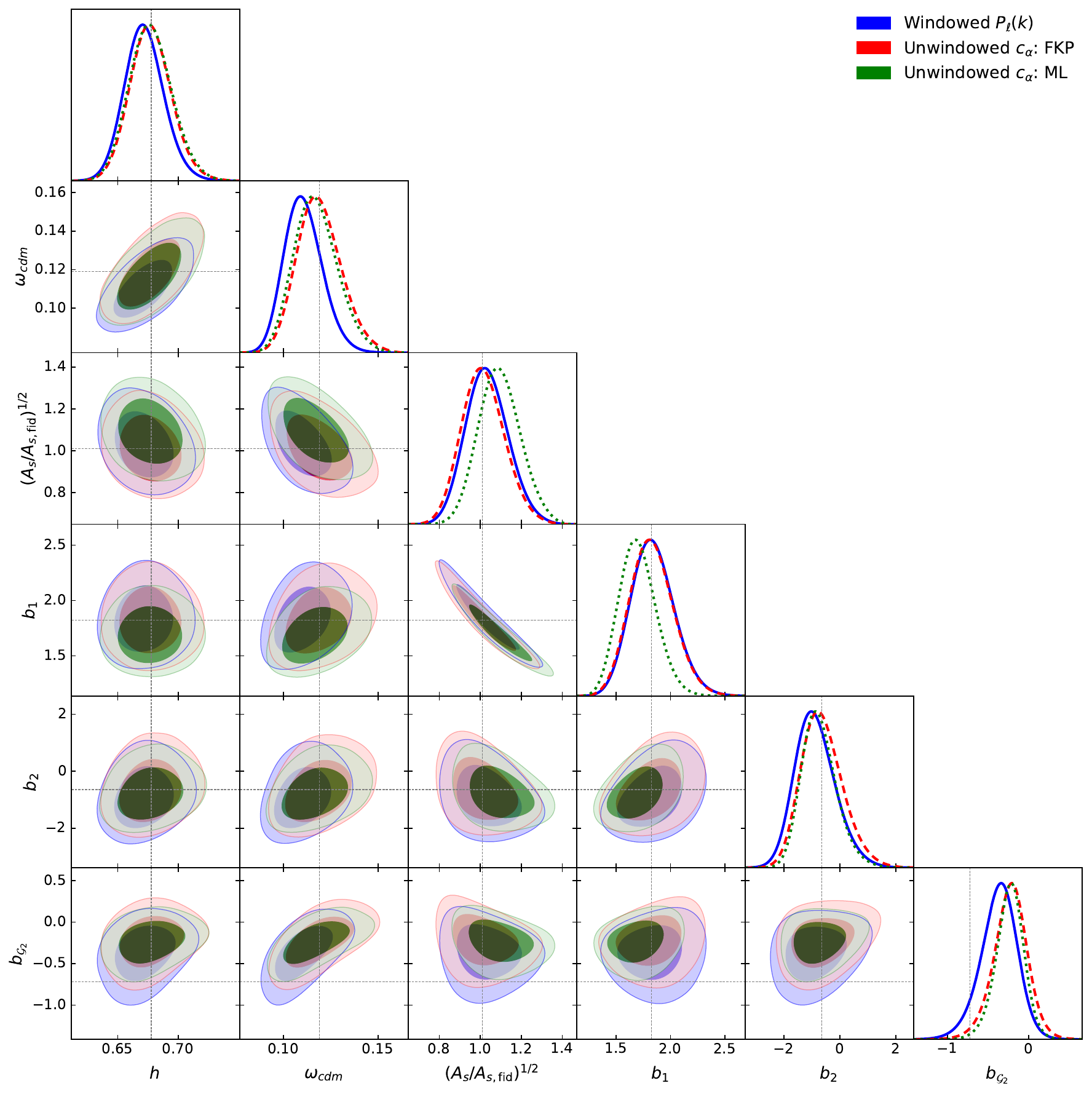}
    \caption{Posterior contours for an MCMC inference on a segment of the BOSS data-set, either analyzing the windowed power spectrum multipoles in 98 bins estimated using \texttt{nbodykit} (full blue) or 15 compressed subspace coefficients estimated via the quadratic estimators discussed in this work, utilizing FKP (dashed red) or ML (dotted green) pixel weights. The latter are analyzed \textit{without} window-convolution of the theory model. 1000 (150) mocks are used to define the sample covariance for the windowed (unwindowed) analyses, supplemented with a correlated theoretical error covariance. We plot only those parameters directly sampled by \texttt{montepython}; another four are marginalized analytically. Thin lines show the parameter values in the fiducial Patchy cosmology, and we set $\log(10^{10}A_{s,\mathrm{fid}}) = 3.044$. Whilst there are some differences between data-sets, all are statistically consistent.}
    \label{fig: param-inference}
\end{figure}

The output parameter contours are displayed in Fig.\,\ref{fig: param-inference}. The results from the three analyses are statistically consistent for all parameters at $\lesssim 0.8\sigma$, which is as expected since all spectra are computed from the same data-set. Furthermore, we find consistent contour widths when analyzing both power spectra and compressed coefficients; whilst Ref.\,\citep{2021PhRvD.103d3508P} demonstrated this for the windowed subspace coefficients, this shows that the same conclusion applies for the directly sampled unwindowed coefficients. The main conclusion from this figure is as follows: estimating subspace coefficients with quadratic estimators gives statistically consistent parameter posteriors to those obtained using the standard power spectrum estimators. When the quadratic estimators are adopted, we do \textit{not} need to convolve the theory model in the likelihood; this expedites the sampling by avoiding the need for Hankel transforms.

Whilst the various data-sets are consistent overall, we do observe small shifts in some parameters, particularly $\omega_{cdm}$ and the primordial amplitude $A_s$. Some variation is expected since (a) the ML and FKP estimators weight the data in slightly different manners, and (b), by throwing away noisy modes, the subspace coefficients have different statistical properties to the data. \resub{Furthermore, Ref.\,\citep{2021PhRvD.103d3508P} demonstrated that noise in the covariance matrix for the 1000-mock windowed power spectrum data-set will cause shifts in the best-fit parameters, whilst this is reduced for the compressed statistic.} If one had analyzed the mean of a set of mocks, these shifts are be expected to vanish \citep[e.g.,][]{2021PhRvD.103d3508P}. Note also that $A_s$ and the linear \resub{galaxy} bias $b_1$ are highly correlated (since the real-space linear power spectrum is proportional to $b_1A_s^2$), thus the $A_s$ shift in the ML data-set is reflected by a corresponding $b_1$ shift. Furthermore, the posteriors are not simply reproducing the fiducial cosmology, as evidenced by the tidal \resub{galaxy} bias $b_{\mathcal{G}_2}$ which is significantly above the input value in all cases. As in previous sections, we find highly consistent results when the subspace coefficients are estimated using FKP and ML pixel weights. In this context, we therefore find little benefit of applying the ML weights instead of the FKP quadratic estimator (though it allows one to compute the Fisher matrix and bias terms on an even footing and is, in practice, not much slower), but we caution that this will depend on the data-set and analysis in question. If one is interested in parameters that impact only large-scale modes (e.g., $f_\mathrm{NL}$), the ML approach is likely to have greater utility as the FKP approach is suboptimal at low-$k$. For this survey, restricting to $k\in[0.01,0.1]\hMpc$ affords similar conclusions, as discussed in Appendix \ref{appen: pix-options}. 

\subsection{Comparison of Quadratic Estimators: ML versus FKP Weights}\label{subsec: ml-fkp-comparision}
The above results merit some discussion regarding the optimality of the ML estimator compared to the quadratic estimator with FKP weights.\footnote{A similar discussion of this in the CMB context is presented in Ref.\,\citep{2004MNRAS.349..603E}.} For a BOSS-like survey, the ML estimator clearly does not lead to a higher-precision power spectrum, nor to a significant increase in constraining power on the underlying parameters. To understand the reasons for this, we first make a number of notes:
\begin{itemize}
    \item If the unclustered number density $n(\vr)$ is uniform, the FKP weighting scheme is optimal, provided the bins are small.\footnote{For finite sized bins, the ratio scales as $\int_{\vk} \left[P(k)+1/\bar{n}\right]^2 \int_{\vk} \left[P(k)+1/\bar{n}\right]^{-2}/N_\mathrm{modes}(k)^2$ in each bin, which is, in practice, very close to unity.} 
    \item On small scales, the FKP estimator is optimal, provided that $P_\fkp$ is close to the associated power spectrum amplitude. This follows directly from the covariance definition \eqref{eq: C-def}; if the window function is compact, $n(\vr)n(\vr')\int_{\vk} e^{i\vk\cdot(\vr-\vr')}P(\vk)\approx P_\fkp n(\vr)\delta_D(\vr-\vr')$, resulting in the FKP form.
    \item If the density of galaxies is low, such that $\bar{n}P(k)\ll 1$, the covariance is dominated by the Poisson shot-noise component, thus ML weights are not important (and the ML and FKP quadratic estimators converge).
    \item Corrections to the FKP weighting are thus expected only to be important on scales where the window function has significant power \textit{and} where $\bar{n}P(k)$ is large.
\end{itemize}

In the above subsections, we have considered the application of the quadratic power spectrum estimators to the power spectrum multipoles of a single BOSS DR12 chunk, which has both large volume ($1.46h^{-3}\mathrm{Gpc}^3$) and relatively low number density ($\bar{n} \sim 2\times 10^{-4}h^3\mathrm{Mpc}^{-3}$). Furthermore, its window function is remarkably compact, as shown in Fig.\,\ref{fig: rand-power-spectrum}. The power spectrum of the random particles (and hence the window function) is less than $1\%$ of its zero lag value even by $k = 0.01\hMpc$, and drops significantly beyond this. In practice, we have omitted the first two $k$-bins ($k<0.01\hMpc$) from all plots in this work, due to their contamination by systematic and integral-constraint effects, \resub{arising from the unknown mean galaxy density and imperfectly subtracted foreground modes.} These results indicate that one may not expect a significant improvement when using the ML weights instead of FKP, matching the conclusions of this work. We argue that it is still useful to implement the quadratic estimator however (even in FKP form), due to the useful property of the resulting spectra being free from the survey window function, and leading order shot-noise and discretization artefacts. Furthermore, the ML approach offers the useful feature that the Fisher matrix and bias terms can be computed directly from the simulation suite used to define the fiducial cosmology, and does not lead to much increased computation time. 
The analysis presented in Appendix \ref{appen: pix-options} provides further proof of this notion, whereupon the fiducial FKP power $P_\fkp$ is reduced by a factor of $10$ to $P_\fkp = 10^{3}h^{-3}\mathrm{Mpc}^3$, considerably below the amplitude of the observed power spectrum multipole at $k=0.25\hMpc$. If the estimated power spectrum multipoles had strong dependence on $P_\fkp$, this would indicate that the survey had some significant anisotropy, and one may expect stronger constraints to be wrought from the ML estimator. In fact, we find very similar power spectrum estimates in this case, suggesting that the impact of the spatially varying number density $n(\vr)$ is small here. 

We may compare the above conclusions to the results of Ref.\,\citep{2019JCAP...09..010C}, which used an optimal quadratic estimator to compute the local-type primordial non-Gaussianity (PNG) parameter $f_\mathrm{NL}^\mathrm{loc}$ from the eBOSS quasar sample over a broad redshift range. The study found $\sim 15-40\%$ improvements in $\sigma(f_\mathrm{NL}^\mathrm{loc})$ (depending on the quasar PNG response), drawing a significantly different conclusion to that of this work. However, this analysis takes a somewhat different approach; whilst we have opted to construct estimators for the galaxy power spectrum itself, Ref.\,\citep{2019JCAP...09..010C} opted to measure $f_\mathrm{NL}^\mathrm{loc}$ directly, and took into account the evolution of the density field across the sample. In this case, the principal difference between the optimal and FKP estimators lies in weighting the galaxies differently as a function of redshift, which, combined with the large redshift range of the sample ($0.8<z<2.2$), leads to the increased precision $f_\mathrm{NL}$ measurements. If one was concerned with measuring just the power spectrum at a single effective redshift and the redshift range was small, we would expect smaller gains, as found here.

\resub{One additional subtlety is worth noting: if one compares the unwindowed FKP-weighted estimators of this work with the standard windowed prescription measuring all power spectrum modes up to some $k_\mathrm{max}$, the latter will generically provide a slightly larger signal-to-noise). This occurs since the window function smooths the spectrum with a characteristic scale $\Delta k$, thus the windowed estimator in fact includes true power spectrum modes up to $k_\mathrm{max}+\Delta k$. Since the variance of a given mode decreases with $k$, this leads to a slight signal-to-noise increase overall, due to the windowed power spectra effectively using a somewhat larger $k_\mathrm{max}$. This effect is not obvious in the results presented herein (since modes at large-$k$ contribute little to the parameter constraints), but would be of greater importance if $k_\mathrm{max}$ were reduced. If one (at least formally) were to force both spectra to use only pre-convolved modes in the same $k$-range, the constraints would be identical however.}

It is instructive to consider the scenarios in which one \textit{would} expect the ML approach to outshine that of FKP. Analytic testing of this is difficult, since the benefits of the ML scheme vanish when employing simplifications such as a fixed number density, and any non-trivial scenario requires inversion of the pixel covariance $\C_\fid$. However, the above discussion implies that the ML approach is expected to be of greatest utility when: (a) the survey volume is small, (b) the number density varies significantly across the survey, with characteristic scale similar to the wavenumbers of interest, (c) the number density is high, such that we are far from the shot-noise dominated regime, or (d) we are interested in parameters depending primarily on large-scale physics, such as $f_\mathrm{NL}$. We leave detailed discussion of these regimes to future work, but note that an example satisfying several of these constraints is the DESI Bright Galaxy Survey \citep{2016arXiv161100036D}.

\begin{figure}
    \centering
    \includegraphics[width=0.5\textwidth]{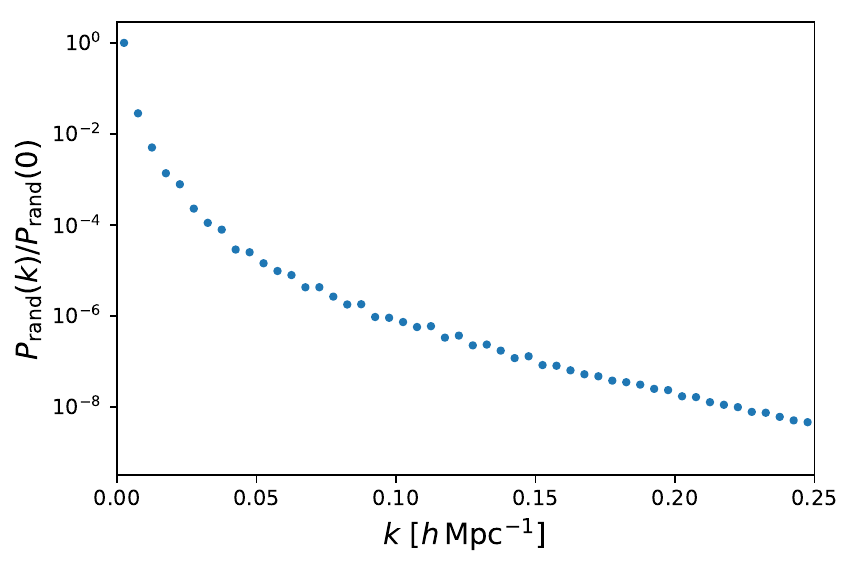}
    \caption{Power spectrum of the unclustered random particles in the BOSS DR12 data chunk considered in this work, normalized to the $k\rightarrow0$ mode. For simplicity, we plot only the monopole, and adopt the same $k$-binning as in previous plots, but including two additional large scale modes at $k<0.01\hMpc$. This is computed as $P_\mathrm{rand}(k_a) = \int_{\vk}\Theta_a(\vk)|n(\vk)|^2/N_\mathrm{modes}^a$, where $N_\mathrm{modes}^a$ is the number of modes in the bin centered at $k_a$ and $n(\vr)$ is the normalized random particle density.}
    \label{fig: rand-power-spectrum}
\end{figure}

\section{Summary and Outlook}\label{sec: conclusion}
Optimal estimation of the galaxy power spectrum was a subject of fervent interest at the turn of the twenty-first century \citep[e.g.,][]{1994ApJ...426...23F,1997ApJ...480...22T,1998PhRvD..57.2117B,2004ApJ...606..702T}, but has thereafter been largely ignored. In this work, we discuss the extent to which quadratic estimators can be used to measure the power spectrum multipoles in a manner that removes the effects of the survey window function. Given a set of simulations with known (or measurable) power spectrum, as well as the galaxy positions and random particle catalog appropriate for the data, quadratic estimators can be constructed to infer the underlying unwindowed power-spectra. Two variants are considered here; a maximum-likelihood (ML) approach (which is optimal in the limit of Gaussian statistics) and a simpler approximation, based on the FKP weighting scheme \citep[e.g.,][]{2017JCAP...07..002H}. Both may be computed from the pixellized density field, assisted by judicious use of Monte Carlo averaging and, for the ML weights, conjugate gradient descent techniques.

Our main conclusions are as follows:
\begin{itemize}
    \item Quadratic estimators allow for robust estimation of unwindowed power spectrum band-powers. Output band-powers have a close-to-diagonal covariance, unlike the windowed case.
    \item The estimators are written as the sum of a smooth fiducial model and a difference between a quantity computed from the data-set and simulations. This implies that they are free from the leading-order effects of gridding, binning, fingers-of-God, and non-Poissonian shot-noise, and allows one to use coarse grids (down to a Nyquist frequency of $k_\mathrm{max}$) in their estimation. \resub{Further more, unless one includes the first $k$-bin, they are free from integral-constraint effects.}
    \item The approach is simply extended to the direct estimation of compressed (unwindowed) power spectrum coefficients. This is fast (since the dimensionality is reduced), and does not require the statistic to be binned.
    \item Applying the ML and FKP quadratic estimators to a data-chunk from BOSS shows them to give statistically compatible parameter constraints to those from conventional approaches. Whilst the ML approach requires a little more computational power and, for BOSS, does not lead to an increase in constraining power, some of the terms in its estimation are somewhat simpler to compute than for FKP weights, and the computation time is very similar in practice. We expect this to be of use for surveys that are particularly anisotropic, small volume, dense or whose analysis principally relies on large-scale modes, such as $f_\mathrm{NL}$-based studies.
\end{itemize}

A number of extensions are possible. Firstly, as discussed in Appendix \ref{appen: non-Gaussian-estimator}, one may formulate analogous \textit{cubic} estimators for the power spectrum that are optimal in the presence of mild non-Gaussianity. Whilst this is unlikely to be of great use for BOSS-like samples, it may be of use for deep surveys such as the DESI Bright Galaxy Survey \citep{2016arXiv161100036D}, which has considerably lower shot-noise and thus contains useful information on short scales. We leave the practical implementation of this algorithm to others, though we note it is only a simple extension of that presented herein. 

Of greater interest is the application to unwindowed bispectra. Via a similar route, one may derive a cubic bispectrum estimator, which is optimal in the limit of mild non-Gaussianity. Unlike for the power spectrum, this estimator does not require a fiducial bispectrum model (and hence knowledge of the bispectrum window), but just a fiducial \textit{power spectrum} model to compute $\Ci_\fid$. In this context, compressing the statistic to a (separable) subspace is of great utility, since (a) it dramatically reduces the number of bispectrum bins, and (b) it obviates the need for the theory model to be binned or window-convolved (which is difficult to do efficiently). This will be considered in future work.

The techniques discussed in this work thus provide a promising renaissance for the optimal estimators of old, allowing for unwindowed spectral estimates to be efficiently computed both in and beyond the Gaussian limit. We expect them to be of great use in future cosmological analyses, particularly for studies based upon the three-point function.

\begin{acknowledgments}
It is a pleasure to thank Giovanni Cabass, \resub{Emanuele Castorina}, Mikhail Ivanov, Marcel Schmittfull, Marko Simonovi\'c, David Spergel and Matias Zaldarriaga for insightful conversations that motivated and improved this work. \resub{We are additionally grateful to the anonymous referees for extensive comments that helped improve the clarity of this work.} OHEP acknowledges funding from the WFIRST program through NNG26PJ30C and NNN12AA01C.

\resub{The authors are pleased to acknowledge that the work reported on in this paper was substantially performed using the Princeton Research Computing resources at Princeton University which is consortium of groups led by the Princeton Institute for Computational Science and Engineering (PICSciE) and Office of Information Technology's Research Computing.}

\end{acknowledgments}

\appendix 

\section{Properties of the Quadratic Estimator}\label{appen: qe-properties}
Below, we demonstrate a number of useful properties of the general quadratic estimator of \eqref{eq: general-qe-est}, and, by extension the ML estimator central to this work. Whilst we denote the parameters by $\vp$, this applies equally well for the subspace coefficients. For clarity, we begin by recapitulating the definition
\beq\label{eq: general-qe}
    \hat{p}_\alpha = p_\alpha^\fid+\sum_\beta F^{-1}_{\alpha\beta}\left(\hat{q}_\beta-\bar{q}_\beta\right), \quad \hat{q}_\alpha = \frac{1}{2}\vd^T\Hi \C_{,\alpha}\Hi\vd
\eeq
(dropping the `QE' superscript for brevity), for arbitrary positive-definite weighting matrix $\mathsf{H}$.

\paragraph{Bias} In expectation, we have
\beq\label{eq: bias1}
    \avEst{\hat{p}_\alpha} = p^\fid_\alpha + \sum_{\beta}F^{-1}_{\alpha\beta}\left(\avEst{\hat{q}_\beta}-\bar{q}_\beta\right) =  p^\fid_\alpha + \frac{1}{2}\sum_{\beta}F^{-1}_{\alpha\beta}\left(\Tr{\C_D\Hi\C_{,\beta}\Hi}-\bar{q}_\beta\right),
\eeq
using the definition $\av{\vd\vd^T}=\C_D$, where $\C_D$ is the data covariance. Assuming the covariance to be linear in the parameter $p_\alpha$, we can write $\C_D-\C_\fid = \sum_{\gamma}\left(p^\mathrm{true}_\gamma-p^\fid_\gamma\right) C_{,\gamma}$ (assuming that the fiducial noise matrix $\N$ to be equal to that of the data and that the sum is over all possible band-powers), 
thus
\beq\label{eq: bias2}
    \avEst{\hat{p}_\alpha} = \sum_\beta F_{\alpha\beta}^{-1} \sum_\gamma p_\gamma^\mathrm{true}  F_{\beta\gamma} = \sum_{\gamma}\delta^K_{\alpha\gamma}\,p_\gamma^\mathrm{true} = p_\alpha^\mathrm{true}, 
\eeq
where we have inserted the bias and Fisher definitions of \eqref{eq: general-qe-est-comp}. 
We thus find the quadratic estimator to be unbiased for all choices of $\mathsf{H}$.

\paragraph{Covariance} The covariance of $\hat{\vp}$ can be written in terms of the covariance of $\hat{\vq}^\mathrm{QE}$;
\beq
    \operatorname{cov}\left(\hat{p}_\alpha, \hat{p}_\beta\right) = \sum_{\gamma\delta}F^{-1}_{\alpha\gamma}F^{-1}_{\beta\delta}\operatorname{cov}\left(\hat{q}_\gamma, \hat{q}_\delta\right),
\eeq
with
\beq
    \operatorname{cov}\left(\hat{q}_\alpha,\hat{q}_\beta\right) = \frac{1}{4}\left[\av{d_id_jd_jd_k}-\av{d_id_j}\av{d_kd_l}\right]Q_\alpha^{ij} Q_\beta^{kl},
\eeq
defining $\left(\Hi\C_{,\alpha}\Hi\right) = \Q_\alpha \equiv -\partial_\alpha \Ci$ and employing the Einstein summation for repeated pixel indices $i,j,k,l$. Employing Wick's theorem,
\beq
    \operatorname{cov}\left(\hat{q}_\alpha,\hat{q}_\beta\right) &=& \frac{1}{4}\left[\C^D_{ik}\C^D_{jl}+\C^D_{jk}\C^D_{il}+\mathsf{T}^D_{ijkl}\right]\Q^{ij}_\alpha \Q^{kl}_\beta\\\nonumber
    &=& \frac{1}{2}\Tr{\C^D\Hi\C_{,\alpha}\Hi\C^D\Hi\C_{,\beta}\Hi} + \frac{1}{4}\mathsf{T}^D_{ijkl}\Q^{ij}_\alpha\Q^{kl}_{\beta}
\eeq
where $\mathsf{T}^D_{ijkl} \equiv \av{d_id_jd_kd_l}_c$ is the connected four-point function of the data. Whilst this is \resub{generally} complex, for the ML estimator with $\vp^\fid = \vp^\mathrm{true}$ and neglecting non-Gaussianity, we have $\C_\fid = \C_D$ and $\mathsf{T}_D = 0$, and thus
\beq\label{eq: ideal-estimator-covariance}
    \operatorname{cov}\left(\hat{q}^\mathrm{ML}_\alpha,\hat{q}^\mathrm{ML}_\beta\right) &\rightarrow& \frac{1}{2}\Tr{\Ci_\fid\C_{,\alpha}\Ci\C_{,\beta}} = F_{\alpha\beta}\\\nonumber
    \Rightarrow \operatorname{cov}\left(\hat{p}^\mathrm{ML}_\alpha,\hat{p}^\mathrm{ML}_\beta\right) &\rightarrow& F_{\alpha\beta}^{-1}
\eeq
thus the Fisher matrix is \resub{just} the inverse covariance matrix. From this, we also expect the quantities $\sum_\beta F_{\alpha\beta}^{-1/2}\hat{q}_\beta^\mathrm{ML}$ to have a covariance close to the identity matrix.\footnote{Ref.\,\citep{2004ApJ...606..702T} advocates plotting this quantity for this reason.}

\paragraph{Optimality}
The Cram\'{e}r-Rao theorem states that, if an estimator $\hat{p}_\alpha$ is optimal, it must satisfy
\beq
    \left.\operatorname{cov}\left(\hat{p}_\alpha,\hat{p}_\beta\right)\right|_\mathrm{CR\,bound} = \av{\frac{1}{2}\frac{\partial^2\mathcal{L}[\vd](\vp)}{\partial p_\alpha\partial p_\beta}}^{-1}\equiv \mathcal{I}_{\alpha\beta}^{-1},
\eeq
where the right-hand-side is the inverse Fisher information for negative log-likelihood $\mathcal{L}[\vd](\vp) = -2\log L[\vd](\vp)$ depending on parameters $\vp$ and data $\vd$ \citep[e.g.][]{2009arXiv0906.0664H}. Assuming the Gaussian likelihood of \eqref{eq: vp-like}, evaluated at the true covariance $\C_D$, the Fisher information is \resub{straightforwardly} derived;
\beq
    \mathcal{I}_{\alpha\beta} &=& \av{\frac{1}{2}\Tr{\Ci\C_{,\alpha}\Ci\C_{,\beta}\Ci\left(2\vd\vd^T - \C\right)}}_{\mathsf{C} = \mathsf{C}^D}\\\nonumber
    &=& \frac{1}{2}\Tr{\Ci\C_{,\alpha}\Ci\C_{,\beta}}_{\C^D = \C} \equiv \left.F_{\alpha\beta}\right|_{\C = \C^D}.
\eeq
Comparison with \eqref{eq: ideal-estimator-covariance} shows that the ML estimator of \eqref{eq: quadratic-estimator-def} saturates its Cram\'er-Rao bound and is thus optimal if $\vp^\mathrm{fid} = \vp^\mathrm{true}$, neglecting non-Gaussianity. This is in fact guaranteed for any ML estimator.

\section{Optimal Estimators with Weak Non-Gaussianity}\label{appen: non-Gaussian-estimator}
If the likelihood of $\vd$ is non-Gaussian, the quadratic power spectrum estimator becomes non-optimal, even in the limit of $\C_\fid = \C_D$. Here, we discuss the leading-order corrections to the maximum likelihood estimator in the presence of weak non-Gaussianity, either from higher-order Poissonian noise corrections or density field non-Gaussianities. We begin with the Edgeworth expansion for the likelihood around the Gaussian form of \eqref{eq: vp-like} (henceforth denoted $L_G$);
\beq\label{eq: edgeworth-expansion}
    L[\vd](\vp) = L_G[\vd](\vp)\left[1 + \frac{1}{3!}\mathsf{B}^{ijk}\mathcal{H}_{ijk} + \frac{1}{4!}\mathsf{T}^{ijkl}\mathcal{H}_{ijkl} + \frac{1}{6!}\left(\mathsf{B}^{ijk}\mathsf{B}^{lmn}+\mathrm{9\,perms.}\right)\mathcal{H}_{ijklmn}+...\right],
\eeq
\citep[e.g.,][]{2005PhRvD..72d3003B,2017arXiv170903452S} where the Hermite tensors $\mathcal{H}_{i_1...i_n}$ are defined in Ref.\,\citep{2017arXiv170903452S}, depending on products of $\Ci \vd$ and $\Ci$.\footnote{\resub{For clarity, we omit the explicit dependence of $L[\vd]$ on the correlators beyond $\vp$. For the purposes of this section, we also omit the subscripts on $\C$, assuming all covariances to be defined at the true parameters.}} Of importance here is the first, defined by\
\beq
    \mathcal{H}_{ijk} = h_ih_jh_k - \left(\C_{ij}h_k + \text{2 perms.}\right),
\eeq
where $h_i \equiv [\Ci\vd]_i$. In \eqref{eq: edgeworth-expansion}, we have included only terms involving the three-point correlator $\mathsf{B}^{ijk}\equiv \av{d^id^jd^k}$ and four-point correlator $\mathsf{T}^{ijkl} \equiv \av{d^id^jd^kd^l} - \left[\av{d^id^j}\av{d^kd^l} + \mathrm{2\,perms.}\right]$. Note that all terms are assumed to be evaluated at the fiducial cosmology. More useful is the negative log-likelihood;
\beq\label{eq: edgeworth-log-expansion}
    \ell[\vd](\vp) = \ell_G[\vd](\vp) - \frac{1}{3!}\mathsf{B}^{ijk}\mathcal{H}_{ijk} - \frac{1}{4!}\mathsf{T}^{ijkl}\mathcal{H}_{ijkl} + \frac{1}{72}\mathsf{B}^{ijk}\mathsf{B}^{lmn}\left[\mathcal{H}_{ijk}\mathcal{H}_{lmn}-\mathcal{H}_{ijklmn}\right]+...,
\eeq
expanding the logarithm to second order and noting that the Hermite tensors are fully symmetric.

To proceed, we require the gradient of $\ell[\vd](\vp)$ with respect to the band-powers, as in \eqref{eq: vp-like-approx}. Note that the band-powers appear \textit{both} in the inverse covariance matrices $\Ci$ (present in the Hermite tensors) and the three-point averages $\B^{ijk}$, through Poisson noise corrections. Following some algebra, this give the leading-order contribution
\beq
    -\partial_\alpha\ell[\vd](\vp) &=& \frac{1}{2}\Q^\alpha_{ij}\left[d^id^j-\C^{ij}\right]+\frac{1}{6}\B^{ijk}_{,\alpha}\Ci_{il}\Ci_{jm}\Ci_{kn}\left[d^ld^md^n-3\C^{lm}d^n\right]\\\nonumber
    &&\,-\frac{1}{2}\B^{ijk}\Ci_{il}\Ci_{jm}\Q^\alpha_{kn}\left[d^ld^md^n-d^l\C^{mn}-\C^{lm}d^n\right]+...,
\eeq
using $\Q_\alpha \equiv \Ci\C_{,\alpha}\Ci$, as before. Here, the first term is the Gaussian piece, with the second and third being cubic corrections from noise and gravitational non-Gaussianities. This motivates the following power spectrum estimator containing quadratic and cubic contributions (by analogy to \eqref{eq: quadratic-estimator-def} for the Gaussian ML case)
\beq
    \hat{p}_\alpha^\mathrm{NG} &=& p_\alpha^\mathrm{fid} +  \sum_{\beta}F^{-1,\mathrm{NG}}_{\alpha\beta}\left(\hat{q}_\beta^\mathrm{NG}-\bar{q}_\beta^\mathrm{NG}\right)\\\nonumber
    \hat{q}_\alpha^\mathrm{NG} &=& \frac{1}{2}\Q^\alpha_{ij}d^id^j +\frac{1}{6}\B^{ijk}_{,\alpha}\Ci_{il}\Ci_{jm}\Ci_{kn}\left[d^ld^md^n-3\C^{lm}d^n\right]\\\nonumber
    &&\,-\frac{1}{2}\B^{ijk}\Ci_{il}\Ci_{jm}\Q^\alpha_{kn}\left[d^ld^md^n-d^l\C^{mn}-\C^{lm}d^n\right].
\eeq
Whilst the bias and Fisher term could be derived from the realization-averaged second derivative $\partial_\alpha\partial_\beta\ell[\vd](\vp)$, this is arduous to compute, so we here take an alternative route, first noting that $\hat{\vq}^\mathrm{NG}$ has expectation
\beq
    \avEst{\hat{q}_\alpha^\mathrm{NG}} &=& \frac{1}{2}\Q^\alpha_{ij}\C^{ij}_D+\frac{1}{6}\B^{ijk}_{,\alpha}\Ci_{il}\Ci_{jm}\Ci_{kn}\B^{lmn}_D-\frac{1}{2}\B^{ijk}\Ci_{il}\Ci_{jm}\Q^\alpha_{kn}\B^{lmn}_D,
\eeq
where the subscripts $D$ denote the data. Both the two- and three-point expectations $\C_D$ and $\B_D$ can be written in the form $\mathsf{X}_D = \mathsf{X} + \sum_\gamma \left(p_\gamma^\mathrm{true}-p_\gamma^\fid\right)X_{,\gamma}$, assuming the fiducial model to have the correct noise properties.\footnote{This also assumes the fiducial model to have the correct gravitational three-point function since it is contained within $\B$; small deviations are unlikely to cause significant bias to the estimator however.} This gives
\beq
    \avEst{\hat{q}_\alpha^\mathrm{NG}} &=& \sum_\gamma \left(p^\mathrm{true}_\gamma-p^\mathrm{fid}_\gamma\right)\left[\frac{1}{2}\Q^\alpha_{ij}\C_{,\gamma}^{ij}+\frac{1}{6}\B_{,\alpha}^{ijk}\Ci_{il}\Ci_{jm}\Ci_{kn}\B^{lmn}_{,\gamma}-\frac{1}{2}\B^{ijk}\Ci_{il}\Ci_{jm}\Q^{\alpha}_{kn}\B_{,\gamma}^{lmn}\right]\\\nonumber
    &&+\left[\frac{1}{2}\Q^\alpha_{ij}\C^{ij}+\frac{1}{6}\B_{,\alpha}^{ijk}\Ci_{il}\Ci_{jm}\Ci_{kn}\B^{lmn}-\frac{1}{2}\B^{ijk}\Ci_{il}\Ci_{jm}\Q^{\alpha}_{kn}\B^{lmn}\right]\\\nonumber
    &\equiv& \sum_\gamma \left(p^\mathrm{true}_\gamma-p^\fid_\gamma\right) F^\mathrm{NG}_{\alpha\gamma} + \bar{q}_\alpha^\mathrm{NG},
\eeq
defining the bias and Fisher matrix in the final line, such that the resulting $\hat{p}_\alpha^\mathrm{NG}$ estimators satisfy $\avEst{\hat{\vp}^\mathrm{NG}} = \vp^\fid$. For $\vp^\mathrm{fid} \rightarrow \vp^\mathrm{true}$, this gives the optimal cubic correction to the power spectrum estimator, to the extent in which we can neglect non-Gaussianity of the signal and noise beyond the three-point function. 

Computation of the relevant terms in $\hat{p}_\alpha$ is possible given the vectors $\Ci\vd$ and $\Q^\alpha\vd$, just as for the Gaussian case. Defining $h_i=[\Ci\vd]_i$ and $\phi_i^\alpha = [\Q^\alpha\vd]_i$, we can write the estimator more succinctly as
\beq
    \hat{q}_\alpha^\mathrm{NG} &=& \frac{1}{2}d^i\phi_i^\alpha + \frac{1}{6}\B_{,\alpha}^{ijk}\left[h_ih_jh_k-3\C_{ij}h_k\right]-\frac{1}{2}\B^{ijk}\left[h_ih_j\phi^\alpha_k-h_i\Q^\alpha_{jk}-\Ci_{ij}\phi^\alpha_k\right]\\\nonumber
    &=&\frac{1}{2}d^i\phi_i^\alpha + \frac{1}{6}\B_{,\alpha}^{ijk}\left[h_ih_jh_k-3\av{\tilde{h}_i\tilde{h}_j}h_k\right]-\frac{1}{2}\B^{ijk}\left[h_ih_j\phi^\alpha_k-h_i\av{\tilde\phi^\alpha_j \tilde{h}_k}-\av{\tilde{h}_i\tilde{h}_j}\phi^\alpha_k\right],
\eeq
where in the second line, we have replaced the $\C_{ij}$ and $\Q^{\alpha}_{ij}$ matrices with averages over Monte Carlo realizations $\tilde{\vd}$ generated at the fiducial cosmology (denoting $\tilde{h}\equiv \Ci\tilde{\vd}$, $\tilde\phi^\alpha\equiv\Q^{\alpha}\tilde{\vec d}$), allowing the operator to be computed as a product of two or three fields in configuration space, given the $\B$ and $\B_{,\alpha}$ weighting tensors, just as in Sec.\,\ref{sec: qe-implementation}. Note the appearance of terms involving \textit{both} Monte Carlo simulations $\vm$ and the data $\vd$. These did not appear in the Gaussian estimator, and require two pixel grids being in memory simultaneously. Whilst on its own this does not pose a difficulty, we caution that, unless carefully treated, it may require extensive computation time to compute a sample covariance matrix, since the term is different for each simulation entering the covariance.

Following a similar approach, we can compute $\bar{q}_\alpha$ via Monte Carlo averaging as in \eqref{eq: bias-fish-as-mc};
\beq
    \bar{q}_\alpha^\mathrm{NG} &=& \frac{1}{2}\av{\tilde{d}^i\tilde{\phi}_i^\alpha} + \frac{1}{6}\av{\B^{ijk}_{,\alpha}\tilde{h}_i\tilde{h}_j\tilde{h}_k} - \frac{1}{2}\av{\B^{ijk}\tilde{h}_i\tilde{h}_j\tilde{\phi}_k^\alpha}.
\eeq
Whilst a similar procedure can be performed for $F_{\alpha\beta}^\mathrm{NG}$, it is more difficult to obtain a simply applicable form in this case, though Ref.\,\citep{2011MNRAS.417....2S} presents a discussion of this in the context of CMB bispectrum estimators. An alternative method is to note that, assuming the estimator to be close-to-optimal,
\beq
    \operatorname{cov}\left(\hat{q}^\mathrm{NG}_\alpha-\hat{q}^\mathrm{G}_\alpha,\hat{q}^\mathrm{NG}_\beta-\hat{q}^\mathrm{G}_\beta\right)\approx F^\mathrm{NG}_{\alpha\beta}-F^\mathrm{G}_{\alpha\beta},
\eeq
where the superscript $G$ indicates the Gaussian results of Sec.\,\ref{sec: qe-theory}. Since the Gaussian matrix can be straightforwardly computed as in Sec.\,\ref{subsec: implement-bias-fish}, computing the non-Gaussian residual in this manner is expected to yield an accurate (and invertible) overall result, assuming it to be a small perturbation.

It remains to specify the form of $\B^{ijk}$. For the spectroscopic surveys considered in this work, we have the three-point average
\beq\label{eq: 3-point-cumulant}
    \B^{ijk} &=& \av{\left[n_g-n\right](\vr_i)\left[n_g-n\right](\vr_j)\left[n_g-n\right](\vr_k)}\\\nonumber
    &=& n(\vr_i)\delta_D(\vr_i-\vr_j)\delta_D(\vr_i-\vr_k) + \left[n(\vr_i)n(\vr_j)\xi(\vr_i-\vr_j)\delta_D(\vr_i-\vr_k) + \text{2\,perms.}\right]\\\nonumber
    &&\,+n(\vr_i)n(\vr_j)n(\vr_k)\zeta(\vr_i,\vr_j,\vr_k),
\eeq
where $\zeta$ is the three-point correlation function. Expressing the correlators in Fourier space, this yields
\beq\label{eq: Bijk-def-Fourier}
    \B^{ijk} &=&  n(\vr_i)\delta_D(\vr_i-\vr_j)\delta_D(\vr_i-\vr_k) + \left[n(\vr_i)n(\vr_j)\delta_D(\vr_i-\vr_k)\int_{\vk}P(\vk;\vr_j)e^{i\vk\cdot(\vr_i-\vr_j)} + \text{2\,perms.}\right]\\\nonumber
    &&\,+n(\vr_i)n(\vr_j)n(\vr_k)\int_{\vk_1\vk_2\vk_3}B(\vk_1,\vk_2,\vk_3)e^{i\vk_1\cdot\vr_{ij}+i\vk_2\cdot\vr_{jk}+i\vk_3\cdot\vr_{ki}}\delD{\vk_1+\vk_2+\vk_3},
\eeq
denoting $\vr_{ab}\equiv \vr_a-\vr_b$ and allowing for the line-of-sight dependence of $P(\vk)$ but assuming the bispectrum $B$ to be isotropic for simplicity.\footnote{This procedure may be analogously generalized for the redshift-space bispectrum multipoles, just involving additional spherical harmonic factors in the estimators.} This now depends on the fiducial power spectrum $P(\vk)$ and bispectrum $B(\vk_1,\vk_2,\vk_3)$. The band-power derivative is \resub{given by}
\beq
    \B_{,\alpha}^{ijk} &=& n(\vr_i)n(\vr_j)\delta_D(\vr_i-\vr_k)\int_{\vk}\Theta_a(\vk)L_\ell(\hat{\vk}\cdot\hat{\vr}_j) e^{i\vk\cdot(\vr_i-\vr_j)} + \text{2\,perms.},
\eeq
arising only from Poisson noise contractions. 

In the case of vanishing gravitational non-Gaussianity, \textit{i.e.} $B(\vk_1,\vk_2,\vk_3)\equiv 0$, the power spectrum estimators simplify, since they depend only on fields at two physical locations rather than three. In particular, the action of $\B$ and $\B_{,\alpha}$ on three fields $\{\vx,\vec{y},\vec{z}\}$ is given by
\beq
    \B^{ijk}x_iy_jz_k &\rightarrow& \int d\vr\,[nxyz](\vr) + \left[\int d\vr_1\,d\vr_2\,[nx](\vr_1)[nyz](\vr_2)\int_{\vk}P(\vk)e^{i\vk\cdot(\vr_1-\vr_2)}+\text{2 perms.}\right]\\\nonumber
    &=& \int d\vr\,[nxyz](\vr)+\left[\int_{\vk}P(\vk)[nx](\vk)[nyz](-\vk)+\text{2 perms.}\right]\\\nonumber
    \B^{ijk}_{,\alpha}x_iy_jz_k &\rightarrow& \int_{\vk}\Theta_\alpha(\vk)[nx](\vk)[nyz](-\vk) + \text{2 perms.}
\eeq
assuming isotropic $P(\vk)$ for simplicity. Both may be \resub{straightforwardly} computed via Fourier transforms and summations of the pixel grid, allowing for the $\hat{q}_\alpha$ coefficients to be swiftly computed. Indeed, the computational expense of computing the estimator at next-to-leading-order in non-Gaussian noise is not parametrically larger than that of the Gaussian case, since it requires only quantities ($h$ and $\phi$) already computed for the Gaussian ML estimator. When allowing for \textit{gravitational} non-Gaussianity, the operators become more difficult to apply, since the action of $\B$ cannot be simply reduced to a single summation in real- or Fourier-space. The calculation is greatly simplified if the fiducial gravitational bispectrum is separable however.

\section{Estimator Simplifications for FKP Pixel Weights}\label{appen: fkp-simplif}
Using the FKP weights of \eqref{eq: H-fkp-def}, the quadratic estimator \eqref{eq: general-qe-est} can be implemented either as for the ML estimator (Sec.\,\ref{sec: qe-implementation}) or more directly, without the use of a fiducial cosmology. In this section, we briefly discuss the latter option.

Writing $\mathsf{H}_\fkp = w_\fkp(\vr)/n(\vr)$ where $w_\fkp(\vr)\equiv \left[1+n(\vr)P_\fkp\right]^{-1}$ is the familiar FKP weight, the action of $\Hi_\fkp$ on a map $x$ is simply
\beq
    \Hi_\fkp[\vx](\vr) = \frac{w_\fkp(\vr)x(\vr)}{n(\vr)},
\eeq
noting that we apply the weights to the grid rather than to the particles directly, in contrary to the windowed FKP approach. $\hat{\vq}^\fkp$ is given by
\beq
    \hat{q}_\alpha^\fkp &=& \frac{1}{2}\vd^T\Hi_\fkp\C_{,\alpha}\Hi_\fkp\vd\\\nonumber
    &=&\frac{1}{2}\int d\vr\,d\vr'\frac{w_\fkp(\vr)d(\vr)}{n(\vr)}\C_{,\alpha}(\vr,\vr')\frac{w_\fkp(\vr')d_\fkp(\vr')}{n(\vr')}\\\nonumber
    &=& \frac{1}{2}\frac{4\pi}{2\ell+1}\sum_{m=-\ell}^\ell \int_{\vk} \Theta_a(\vk)Y_{\ell m}^*(\vk)[w_\fkp d](-\vk)[Y_{\ell m}w_\fkp d ](\vk),
\eeq
inserting \eqref{eq: C-alpha-def} and applying the Yamamoto approximation, as in \eqref{eq: applyC-alpha}. This \resub{can be easily} evaluated with a Fourier transform. Following the form of \eqref{eq: qe-no-fid}, the bias and Fisher components are given similarly as
\beq
    \bar{q}^\fkp_\alpha &=& \frac{1}{2}\Tr{\Hi_\fkp \C_{,\alpha}\Hi_\fkp\N} = \frac{(1+\alpha)}{2}\int d\vr\,n(\vr)w_\fkp^2(\vr)\int_{\vk}\Theta_a(\vk)L_\ell(\hat{\vk}\cdot\hat{\vr})\\\nonumber
    &=&\frac{(1+\alpha)}{2}\frac{4\pi}{2\ell+1}\sum_{m=-\ell}^\ell\int d\vr\,n(\vr)w^2_\fkp(\vr)Y_{\ell m}(\vr)\int_{\vk}Y_{\ell m}^*(\vk)\Theta_a(\vk)\\\nonumber
    F_{\alpha\beta}^\fkp &=& \frac{1}{2}\Tr{\Hi_\fkp \C_{,\alpha}\Hi_\fkp\C_{,\beta}}\\\nonumber
    &=& \frac{1}{2}\int d\vr\,d\vr'\,n(\vr)w_\fkp(\vr)n(\vr')w_\fkp(\vr')\int_{\vk\,\vk'}\Theta_a(\vk)\Theta_b(\vk)L_\ell(\hat{\vk}\cdot\hat{\vr}')L_{\ell'}(\hat{\vk}'\cdot\hat{\vr})e^{i(\vk-\vk')\cdot(\vr-\vr')}\\\nonumber
    &=& \frac{1}{2}\frac{(4\pi)^2}{(2\ell+1)(2\ell'+1)}\sum_{m=-\ell}^\ell \sum_{m'=-\ell'}^{\ell'}\int_{\vk\,\vk'}\ft{w_\fkp n Y_{\ell'm'}}(\vk-\vk')\ft{w_\fkp n Y_{\ell m}}(\vk'-\vk)Y^*_{\ell m}(\vk)\Theta_a(\vk)Y^*_{\ell'm'}(\vk')\Theta_b(\vk').
\eeq
The Fisher matrix term can be written as a convolution and a Fourier-space summation allowing for straightforward computation. Ignoring the effects of \resub{pixellization}, $\bar{q}_\alpha = \tfrac{1}{2}(1+\alpha) N_\mathrm{modes}^a\delta^K_{\ell 0}\int d\vr\,n(\vr)w^2_\fkp(\vr)$ where $N_\mathrm{modes}^a = \int d\vk\,\Theta_a(\vk)$, and $\hat{q}_\alpha$ and $\bar{q}_\alpha$ are just the usual windowed FKP spectrum and shot-noise multiplied by $N_\mathrm{modes}^a$ and a constant factor. In our case, multiplication by the Fisher matrix inverse $F^{-1,\mathrm{FKP}}_{\alpha\beta}$ acts to remove the effects of the window function and the bias term removes Poissonian shot-noise. For a uniform survey and isotropic $P(\vk)$, $\ft{w_\fkp n}(\vp) = \delD{\vp}w_\fkp n$, thus $F_{\alpha\beta}^\fkp = \frac{1}{2}\delta^K_{\alpha\beta}N_\mathrm{modes}^\alpha w_\fkp^2n^2$, such that all bins are uncorrelated and the density field enters only via the combination $|d(\vk)|^2$, as expected.

\section{Quadratic Estimator Hyperparameters}\label{appen: pix-options}
Below, we provide brief discussion on of the effects of varying a number of hyperparameters controlling the implementation and application of the quadratic estimators, including the effects of binning, weighting and \resub{pixellization}.

\paragraph{Size of \resub{Pixellization} Grid}
A necessary step for computation of the quadratic estimators is assigning the data to a cuboidal grid, allowing for Fourier transforms to be easily computed via the FFT algorithm. For the simple windowed power spectrum estimates, the pixel-width $L$ is usually chosen to ensure that the associated Nyquist frequency ($k_\mathrm{Nyq} = 2\pi/L$) is at least twice $k_\mathrm{max}$; this ensures that the spectra are free from discretization artefacts, though, due to the $\mathcal{O}(N_g\log N_g)$ scaling of FFTs, a denser grid requires greater computational time. In the quadratic estimator formalism, the leading-order effects of \resub{pixellization} are cancelled since the band-powers are computed from a difference of quantities computed in the data and simulations; it may thus be feasible to use a smaller grid-size. 

\begin{figure}
\centering
\begin{minipage}[t]{.48\textwidth}
  \centering
  \includegraphics[width=.95\linewidth]{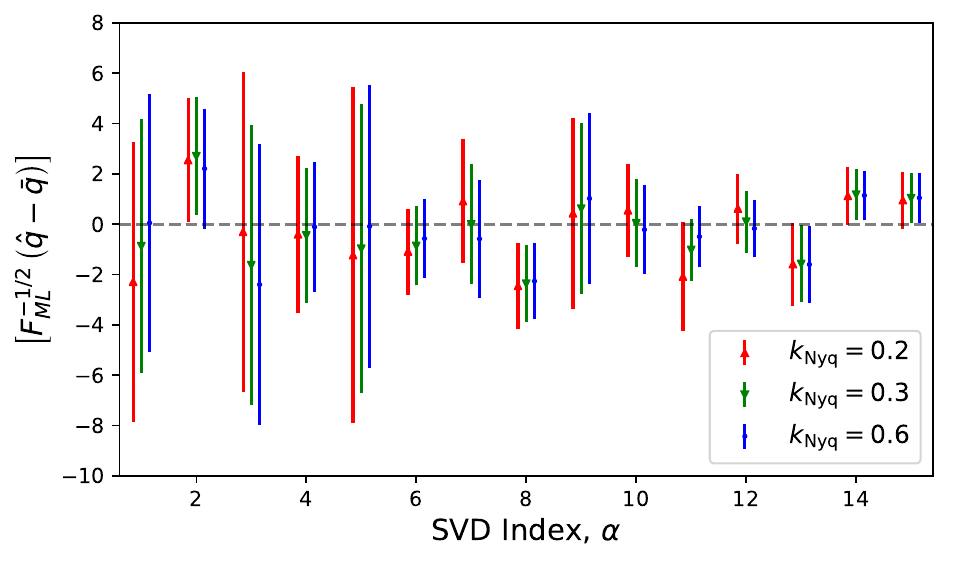}
  \caption{Comparison of the subspace coefficients estimated from BOSS DR12 using three different choices of \resub{pixellization} grid, with associated Nyquist frequencies given in  $\hMpc$ units in the caption. The quadratic estimator is rescaled as in Fig.\,\ref{fig: subspace-coeff} (right panel) for clarity, with the dashed line indicating the values in the fiducial cosmology. All points are computed using the ML quadratic estimator of \eqref{eq: quadratic-estimator-def}, using 150 Patchy mocks to compute the covariances. We note very little difference from reducing the pixel density.}
  \label{fig: sv-grid-comparison}
\end{minipage}%
\quad
\begin{minipage}[t]{.48\textwidth}
  \centering
  \includegraphics[width=.95\linewidth]{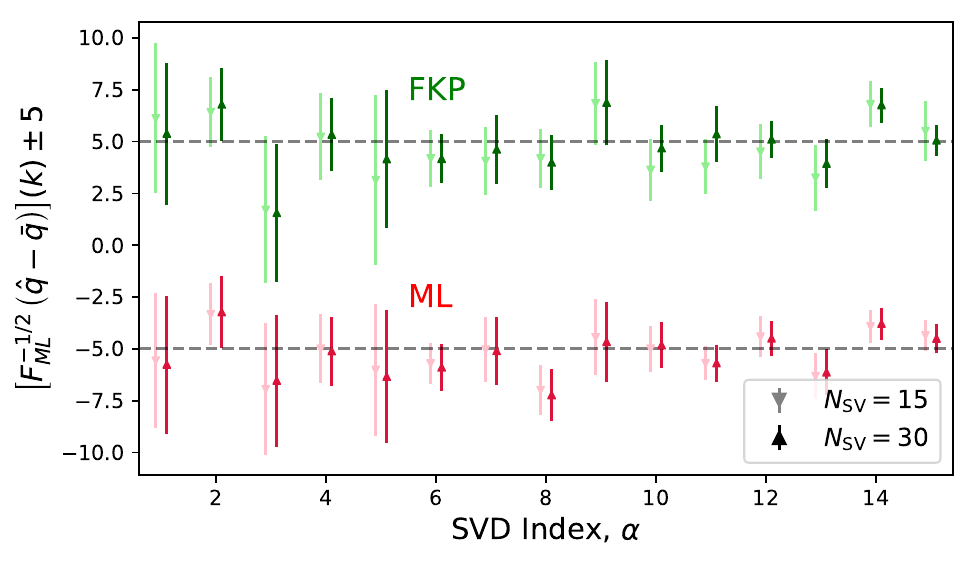}
  \caption{Estimates of the subspace coefficients as a function of total number of coefficients sampled ($N_\mathrm{SV}$). Results with 15 (30) coefficients are shown in light (dark) with the top (bottom) data-sets computed using the FKP (ML) estimator. All cases assume $k_\mathrm{Nyq} = 0.3\hMpc$, with the variances computed from 150 Patchy mocks, and data \resub{are} rescaled as in Fig.\,\ref{fig: subspace-coeff}. Since the coefficients are somewhat correlated, there is a slight decrease in the errorbars as $N_\mathrm{SV}$ is increased when using FKP weights.}
  \label{fig: sv-N-sv-comparison}
\end{minipage}
\end{figure}

Subspace coefficients computed with the ML estimator (Sec.\,\ref{sec: subspace-qe}) with three choices of pixel-width are shown in Fig.\,\ref{fig: sv-grid-comparison}. We observe consistent results with all three, albeit with a very slight increase in the error-bars for $k_\mathrm{Nyq} = 0.2\hMpc$. This is particularly notable for the coarsest grid, since $k_\mathrm{Nyq}$ is below the $k_\mathrm{max} = 0.25\hMpc$ used to construct the basis vectors. This conclusion depends somewhat on the choice of basis; the subspace vectors preferentially weight wavenumbers most sensitive to cosmological and nuisance parameters, thus the high-$k$ regions most affected by gridding artefacts are downweighted. Repeating the analysis for the band-powers shows that the results are independent of $k_\mathrm{Nyq}$ providing $k_\mathrm{Nyq}>k_\mathrm{max}$. The latter conclusion applies also to the FKP-based quadratic estimator. This weak restriction allows us to use relatively coarse grids and hence expedite the computation.

\paragraph{Number of Basis Vectors}
As discussed in Sec.\,\ref{subsec: sv-qe}, the measured subspace coefficients carry some dependence on the total number estimated, $N_\mathrm{SV}$. This is shown in Fig.\,\ref{fig: sv-N-sv-comparison}, plotting the first 15 subspace coefficients from the ML and FKP estimators using two choices of $N_\mathrm{SV}$. For the ML case, the results are consistent, but the FKP approach suffers from inflated parameter covariances with $N_\mathrm{SV} = 15$, particularly for the later coefficients. This arises due to weighting-specific correlations between measured and unmeasured coefficients that bias the estimator (and increase its variance). We conclude that, when using the FKP estimator,  it is important to measure slightly more coefficients than will be used in the final analysis.

\begin{figure}
\centering
\begin{minipage}[t]{.4\textwidth}
  \centering
    \includegraphics[width=0.95\textwidth]{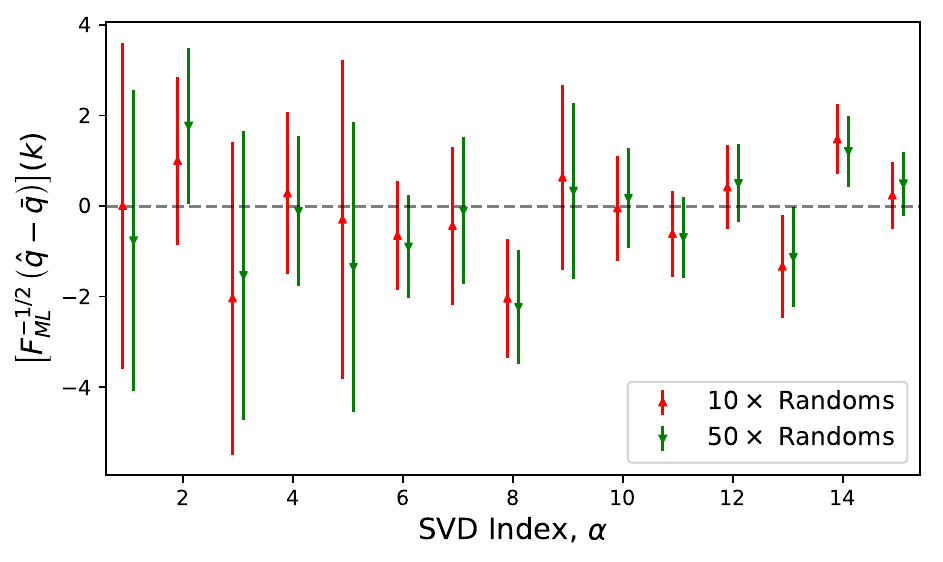}
    \caption{As Fig.\,\ref{fig: sv-grid-comparison}, but comparing the estimated subspace coefficients when the number of random particles is reduced. Results from random catalogs with $10\times$ ($50\times$) the galaxy density are shown in red (green) and all data-sets use $k_\mathrm{Nyq} = 0.3\hMpc$, the ML estimator, and $N_\mathrm{SV} = 30$. Whilst there are small shifts in the coefficients, these are not systematic, and are well within the error-bars.}
    \label{fig: sv-N-rand-comparison}
\end{minipage}%
\quad
\begin{minipage}{.58\textwidth}
  \centering
  \includegraphics[width=0.7\textwidth]{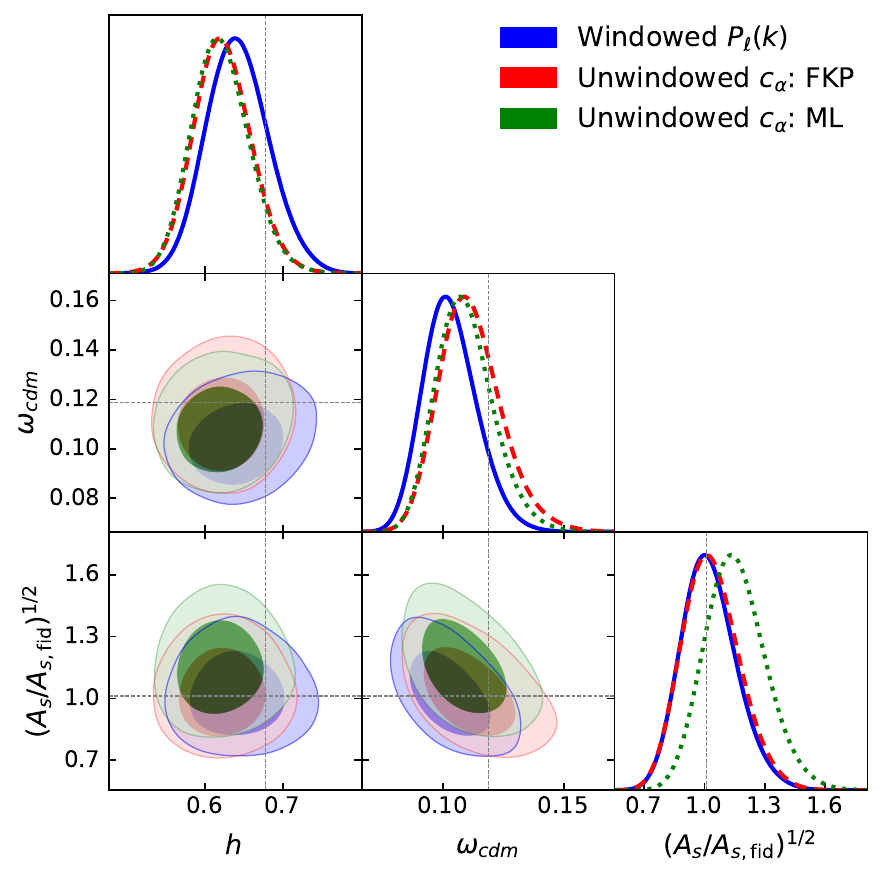}
  \caption{As Fig.\,\ref{fig: param-inference}, but reducing $k_\mathrm{max}$ to $0.1\hMpc$. For clarity, we show only the cosmological parameters. Our conclusions regarding the efficacy of the various estimators are unchanged.}
  \label{fig: param-inference-lowk}
 \end{minipage}
 \end{figure}

\paragraph{Number of Random Particles}
The random particle density $n_r(\vr)$ enters the algorithm both to define the pixellized data $d(\vr) = n_g(\vr)-n_r(\vr)$, and as a proxy of the background number density $n(\vr)$ appearing in the pixel covariance matrix \eqref{eq: C-def}. Fig.\,\ref{fig: sv-N-rand-comparison} considers the impact of this for the ML case (which is expected to be the most affected, since several of the $n(\vr)$ fields cancel when applying FKP weighting). Reducing the density of randoms from 50 to 10 times that of the data causes slight stochastic shifts in the output coefficients which is expected (and a similar effect is seen for the windowed power spectrum estimators). Noting that the shifts are well within the error bars even for $10\times$ randoms, we expect our main results (which use $50\times$ randoms) to be robust.

\paragraph{Choice of $k_\mathrm{max}$}
When performing parameter inference with the quadratic estimators, an important hyperparameter is the maximum wavenumber used, here set to $k_\mathrm{max} = 0.25\hMpc$ in this work. Whilst the exact choice is not of particular importance, due to the correlated theoretical error that reduces sensitivity to short, poorly modelled, scales, by substantially reducing $k_\mathrm{max}$, we can check how the quadratic estimator affects the information content of the large-scale modes. Since much of the information on cosmological parameters comes from intermediate scales (which allow one to constrain nuisance parameters, \citep{2020JCAP...05..042I}), this is a valid test of the approach. To this end, we have reduced $k_\mathrm{max}$ to $0.1\hMpc$, and recomputed the subspace coefficients of Sec.\,\ref{sec: subspace-qe} (since a change in $k_\mathrm{max}$ necessarily leads to a change in the basis vectors), before running the parameter inference as in Sec.\,\ref{sec: boss-analysis}. The resulting constraints on cosmological parameters are shown in Fig.\,\ref{fig: param-inference-lowk}. Notably, our conclusions are very similar to those with $k_\mathrm{max} = 0.25\hMpc$: the quadratic estimators lead to slight (but not significant) shifts in the cosmological parameters, but the FKP and ML weights are similar, though with a slight shift to higher $A_s$ in the latter case. 

\paragraph{FKP Weighting}
Finally, we comment on the effects of changing the FKP coefficient, $P_\fkp$ appearing in the FKP pixel weight matrix of \eqref{eq: H-fkp-def}. By reducing this from $P_\fkp = 10^4h^{-3}\mathrm{Mpc}^{3}$ to $P_\fkp = 10^3h^{-3}\mathrm{Mpc}^{3}$ (significantly below the value of the power spectrum monopole on all tested scales), we would expect to increase the power spectrum error bars if the effects of non-uniform survey number density were significant. The resulting power spectrum multipoles are shown in Fig.\,\ref{fig: pk-fkp-weight-test}, alongside those with the fiducial FKP weight used in Fig.\,\ref{fig: pk-full-plot}. Whilst there are minor perturbations to the spectra due to the different choice of weighting, we find statistically consistent results, and, importantly, do not observe inflated error bars, even at low-$k$. Whilst this may appear somewhat surprising, given that the FKP weight is not optimal in this case, it indicates that, for a broad survey such as BOSS, the exact pixel weighting scheme does not have a significant impact, likely because the window function is relatively compact, and the number density relatively consistent across the survey. That this does not have a significant impact further suggests that the ML approach will not greatly improve the BOSS measurement of $P_\ell(k)$, just as found in this work.

\begin{figure}
    \centering
    \includegraphics[width=0.6\textwidth]{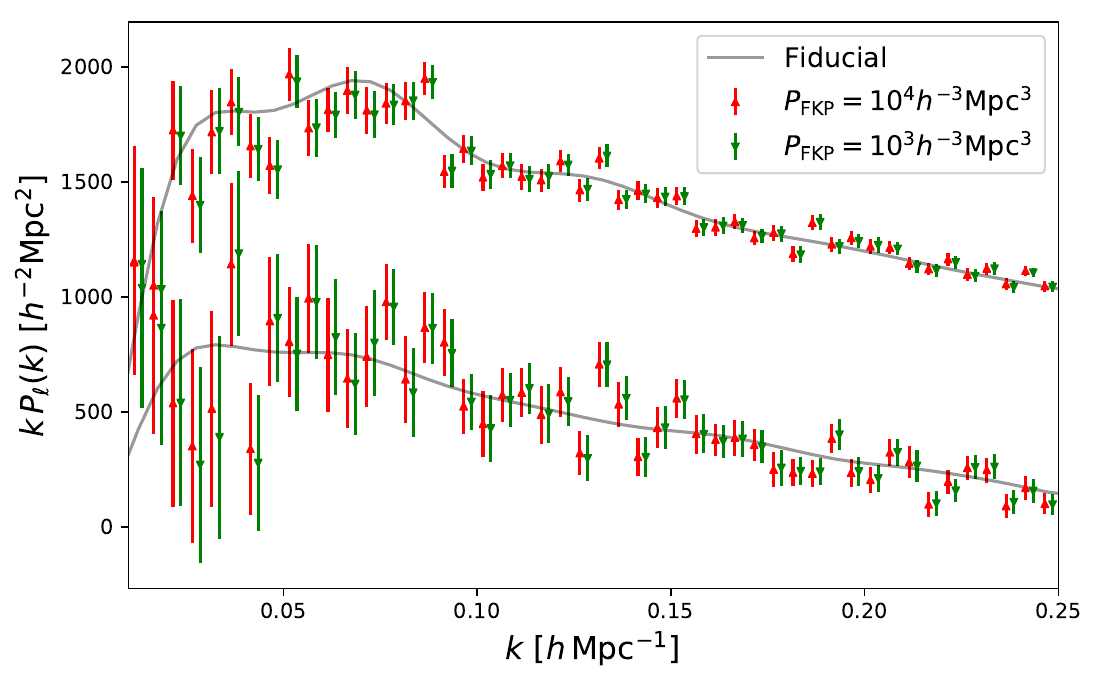}
    \caption{As Fig.\,\ref{fig: pk-full-plot}, but comparing two different FKP weighting schemes, with $P_\fkp= 10^4h^{-3}\mathrm{Mpc}^3$ (red) and $10^3h^{-3}\mathrm{Mpc}^3$ (green). The second choice of weighting is suboptimal, since it significantly underestimates the true power. In both cases, spectra are computed using the quadratic estimators of Sec.\,\ref{sec: qe-theory}, and are thus unbiased by the survey window function.}
    \label{fig: pk-fkp-weight-test}
\end{figure}

\bibliographystyle{JHEP}
\bibliography{adslib,otherlib}

\providecommand{\href}[2]{#2}\begingroup\raggedright\begin{thebibliography}{10}

\bibitem{2017MNRAS.466.2242B}
F.~{Beutler}, H.-J. {Seo}, S.~{Saito}, C.-H. {Chuang}, A.~J. {Cuesta}, D.~J.
  {Eisenstein} et~al., \emph{{The clustering of galaxies in the completed
  SDSS-III Baryon Oscillation Spectroscopic Survey: anisotropic galaxy
  clustering in Fourier space}},
  \href{https://doi.org/10.1093/mnras/stw3298}{\emph{\mnras} {\bfseries 466}
  (2017) 2242} [\href{https://arxiv.org/abs/1607.03150}{{\ttfamily
  1607.03150}}].

\bibitem{2020MNRAS.498.2492G}
H.~{Gil-Mar{\'\i}n}, J.~E. {Bautista}, R.~{Paviot}, M.~{Vargas-Maga{\~n}a},
  S.~{de la Torre}, S.~{Fromenteau} et~al., \emph{{The Completed SDSS-IV
  extended Baryon Oscillation Spectroscopic Survey: measurement of the BAO and
  growth rate of structure of the luminous red galaxy sample from the
  anisotropic power spectrum between redshifts 0.6 and 1.0}},
  \href{https://doi.org/10.1093/mnras/staa2455}{\emph{\mnras} {\bfseries 498}
  (2020) 2492} [\href{https://arxiv.org/abs/2007.08994}{{\ttfamily
  2007.08994}}].

\bibitem{1994ApJ...426...23F}
H.~A. {Feldman}, N.~{Kaiser} and J.~A. {Peacock}, \emph{{Power-Spectrum
  Analysis of Three-dimensional Redshift Surveys}},
  \href{https://doi.org/10.1086/174036}{\emph{\apj} {\bfseries 426} (1994) 23}
  [\href{https://arxiv.org/abs/astro-ph/9304022}{{\ttfamily
  astro-ph/9304022}}].

\bibitem{2003ApJ...595..577Y}
K.~{Yamamoto}, \emph{{Optimal Weighting Scheme in Redshift-Space Power Spectrum
  Analysis and a Prospect for Measuring the Cosmic Equation of State}},
  \href{https://doi.org/10.1086/377488}{\emph{\apj} {\bfseries 595} (2003) 577}
  [\href{https://arxiv.org/abs/astro-ph/0208139}{{\ttfamily
  astro-ph/0208139}}].

\bibitem{2006PASJ...58...93Y}
K.~{Yamamoto}, M.~{Nakamichi}, A.~{Kamino}, B.~A. {Bassett} and H.~{Nishioka},
  \emph{{A Measurement of the Quadrupole Power Spectrum in the Clustering of
  the 2dF QSO Survey}},
  \href{https://doi.org/10.1093/pasj/58.1.93}{\emph{\pasj} {\bfseries 58}
  (2006) 93} [\href{https://arxiv.org/abs/astro-ph/0505115}{{\ttfamily
  astro-ph/0505115}}].

\bibitem{2015MNRAS.453L..11B}
D.~{Bianchi}, H.~{Gil-Mar{\'\i}n}, R.~{Ruggeri} and W.~J. {Percival},
  \emph{{Measuring line-of-sight-dependent Fourier-space clustering using
  FFTs}}, \href{https://doi.org/10.1093/mnrasl/slv090}{\emph{\mnras} {\bfseries
  453} (2015) L11} [\href{https://arxiv.org/abs/1505.05341}{{\ttfamily
  1505.05341}}].

\bibitem{2017JCAP...07..002H}
N.~{Hand}, Y.~{Li}, Z.~{Slepian} and U.~{Seljak}, \emph{{An optimal FFT-based
  anisotropic power spectrum estimator}},
  \href{https://doi.org/10.1088/1475-7516/2017/07/002}{\emph{\jcap} {\bfseries
  2017} (2017) 002} [\href{https://arxiv.org/abs/1704.02357}{{\ttfamily
  1704.02357}}].

\bibitem{2017JCAP...04..029S}
D.~{Sorini}, \emph{{An optimally weighted estimator of the linear power
  spectrum disentangling the growth of density perturbations across galaxy
  surveys}}, \href{https://doi.org/10.1088/1475-7516/2017/04/029}{\emph{\jcap}
  {\bfseries 2017} (2017) 029}
  [\href{https://arxiv.org/abs/1612.00725}{{\ttfamily 1612.00725}}].

\bibitem{1998ApJ...499..555T}
M.~{Tegmark}, A.~J.~S. {Hamilton}, M.~A. {Strauss}, M.~S. {Vogeley} and A.~S.
  {Szalay}, \emph{{Measuring the Galaxy Power Spectrum with Future Redshift
  Surveys}}, \href{https://doi.org/10.1086/305663}{\emph{\apj} {\bfseries 499}
  (1998) 555} [\href{https://arxiv.org/abs/astro-ph/9708020}{{\ttfamily
  astro-ph/9708020}}].

\bibitem{1997PhRvD..55.5895T}
M.~{Tegmark}, \emph{{How to measure CMB power spectra without losing
  information}}, \href{https://doi.org/10.1103/PhysRevD.55.5895}{\emph{\prd}
  {\bfseries 55} (1997) 5895}
  [\href{https://arxiv.org/abs/astro-ph/9611174}{{\ttfamily
  astro-ph/9611174}}].

\bibitem{1997ApJ...480...22T}
M.~{Tegmark}, A.~N. {Taylor} and A.~F. {Heavens}, \emph{{Karhunen-Lo{\`e}ve
  Eigenvalue Problems in Cosmology: How Should We Tackle Large Data Sets?}},
  \href{https://doi.org/10.1086/303939}{\emph{\apj} {\bfseries 480} (1997) 22}
  [\href{https://arxiv.org/abs/astro-ph/9603021}{{\ttfamily
  astro-ph/9603021}}].

\bibitem{1998PhRvD..57.2117B}
J.~R. {Bond}, A.~H. {Jaffe} and L.~{Knox}, \emph{{Estimating the power spectrum
  of the cosmic microwave background}},
  \href{https://doi.org/10.1103/PhysRevD.57.2117}{\emph{\prd} {\bfseries 57}
  (1998) 2117} [\href{https://arxiv.org/abs/astro-ph/9708203}{{\ttfamily
  astro-ph/9708203}}].

\bibitem{1999PhRvD..59b7302B}
J.~{Borrill}, \emph{{Power spectrum estimators for large CMB datasets}},
  \href{https://doi.org/10.1103/PhysRevD.59.027302}{\emph{\prd} {\bfseries 59}
  (1999) 027302} [\href{https://arxiv.org/abs/astro-ph/9712121}{{\ttfamily
  astro-ph/9712121}}].

\bibitem{1999ApJ...510..551O}
S.~P. {Oh}, D.~N. {Spergel} and G.~{Hinshaw}, \emph{{An Efficient Technique to
  Determine the Power Spectrum from Cosmic Microwave Background Sky Maps}},
  \href{https://doi.org/10.1086/306629}{\emph{\apj} {\bfseries 510} (1999) 551}
  [\href{https://arxiv.org/abs/astro-ph/9805339}{{\ttfamily
  astro-ph/9805339}}].

\bibitem{2005astro.ph..3603H}
A.~J.~S. {Hamilton}, \emph{{Power Spectrum Estimation I. Basics}}, {\emph{arXiv
  e-prints} (2005) astro}
  [\href{https://arxiv.org/abs/astro-ph/0503603}{{\ttfamily
  astro-ph/0503603}}].

\bibitem{2005astro.ph..3604H}
A.~J.~S. {Hamilton}, \emph{{Power Spectrum Estimation II. Linear Maximum
  Likelihood}}, {\emph{arXiv e-prints} (2005) astro}
  [\href{https://arxiv.org/abs/astro-ph/0503604}{{\ttfamily
  astro-ph/0503604}}].

\bibitem{2015MNRAS.454.1266S}
R.~E. {Smith} and L.~{Marian}, \emph{{Towards optimal estimation of the galaxy
  power spectrum}}, \href{https://doi.org/10.1093/mnras/stv2042}{\emph{\mnras}
  {\bfseries 454} (2015) 1266}
  [\href{https://arxiv.org/abs/1503.06830}{{\ttfamily 1503.06830}}].

\bibitem{2016MNRAS.457.4285S}
R.~E. {Smith} and L.~{Marian}, \emph{{What is the optimal way to measure the
  galaxy power spectrum?}},
  \href{https://doi.org/10.1093/mnras/stw282}{\emph{\mnras} {\bfseries 457}
  (2016) 4285} [\href{https://arxiv.org/abs/1507.04365}{{\ttfamily
  1507.04365}}].

\bibitem{2019JCAP...09..010C}
E.~{Castorina}, N.~{Hand}, U.~{Seljak}, F.~{Beutler}, C.-H. {Chuang}, C.~{Zhao}
  et~al., \emph{{Redshift-weighted constraints on primordial non-Gaussianity
  from the clustering of the eBOSS DR14 quasars in Fourier space}},
  \href{https://doi.org/10.1088/1475-7516/2019/09/010}{\emph{\jcap} {\bfseries
  2019} (2019) 010} [\href{https://arxiv.org/abs/1904.08859}{{\ttfamily
  1904.08859}}].

\bibitem{2000MNRAS.317L..23H}
A.~J.~S. {Hamilton}, M.~{Tegmark} and N.~{Padmanabhan}, \emph{{Linear redshift
  distortions and power in the IRAS Point Source Catalog Redshift Survey}},
  \href{https://doi.org/10.1046/j.1365-8711.2000.03888.x}{\emph{\mnras}
  {\bfseries 317} (2000) L23}
  [\href{https://arxiv.org/abs/astro-ph/0004334}{{\ttfamily
  astro-ph/0004334}}].

\bibitem{2002MNRAS.335..887T}
M.~{Tegmark}, A.~J.~S. {Hamilton} and Y.~{Xu}, \emph{{The power spectrum of
  galaxies in the 2dF 100k redshift survey}},
  \href{https://doi.org/10.1046/j.1365-8711.2002.05622.x}{\emph{\mnras}
  {\bfseries 335} (2002) 887}
  [\href{https://arxiv.org/abs/astro-ph/0111575}{{\ttfamily
  astro-ph/0111575}}].

\bibitem{2002ApJ...571..191T}
M.~{Tegmark}, S.~{Dodelson}, D.~J. {Eisenstein}, V.~{Narayanan},
  R.~{Scoccimarro}, R.~{Scranton} et~al., \emph{{The Angular Power Spectrum of
  Galaxies from Early Sloan Digital Sky Survey Data}},
  \href{https://doi.org/10.1086/339894}{\emph{\apj} {\bfseries 571} (2002) 191}
  [\href{https://arxiv.org/abs/astro-ph/0107418}{{\ttfamily
  astro-ph/0107418}}].

\bibitem{2004ApJ...606..702T}
M.~{Tegmark}, M.~R. {Blanton}, M.~A. {Strauss}, F.~{Hoyle}, D.~{Schlegel},
  R.~{Scoccimarro} et~al., \emph{{The Three-Dimensional Power Spectrum of
  Galaxies from the Sloan Digital Sky Survey}},
  \href{https://doi.org/10.1086/382125}{\emph{\apj} {\bfseries 606} (2004) 702}
  [\href{https://arxiv.org/abs/astro-ph/0310725}{{\ttfamily
  astro-ph/0310725}}].

\bibitem{2004MNRAS.349..603E}
G.~{Efstathiou}, \emph{{Myths and truths concerning estimation of power
  spectra: the case for a hybrid estimator}},
  \href{https://doi.org/10.1111/j.1365-2966.2004.07530.x}{\emph{\mnras}
  {\bfseries 349} (2004) 603}
  [\href{https://arxiv.org/abs/astro-ph/0307515}{{\ttfamily
  astro-ph/0307515}}].

\bibitem{2017JCAP...12..009S}
U.~{Seljak}, G.~{Aslanyan}, Y.~{Feng} and C.~{Modi}, \emph{{Towards optimal
  extraction of cosmological information from nonlinear data}},
  \href{https://doi.org/10.1088/1475-7516/2017/12/009}{\emph{\jcap} {\bfseries
  2017} (2017) 009} [\href{https://arxiv.org/abs/1706.06645}{{\ttfamily
  1706.06645}}].

\bibitem{2011MNRAS.417....2S}
K.~M. {Smith} and M.~{Zaldarriaga}, \emph{{Algorithms for bispectra:
  forecasting, optimal analysis and simulation}},
  \href{https://doi.org/10.1111/j.1365-2966.2010.18175.x}{\emph{\mnras}
  {\bfseries 417} (2011) 2}
  [\href{https://arxiv.org/abs/astro-ph/0612571}{{\ttfamily
  astro-ph/0612571}}].

\bibitem{2021PhRvD.103d3508P}
O.~H.~E. {Philcox}, M.~M. {Ivanov}, M.~{Zaldarriaga}, M.~{Simonovi{\'c}} and
  M.~{Schmittfull}, \emph{{Fewer mocks and less noise: Reducing the
  dimensionality of cosmological observables with subspace projections}},
  \href{https://doi.org/10.1103/PhysRevD.103.043508}{\emph{\prd} {\bfseries
  103} (2021) 043508} [\href{https://arxiv.org/abs/2009.03311}{{\ttfamily
  2009.03311}}].

\bibitem{2016arXiv161100036D}
{DESI Collaboration}, A.~{Aghamousa}, J.~{Aguilar}, S.~{Ahlen}, S.~{Alam},
  L.~E. {Allen} et~al., \emph{{The DESI Experiment Part I: Science,Targeting,
  and Survey Design}}, {\emph{arXiv e-prints} (2016) arXiv:1611.00036}
  [\href{https://arxiv.org/abs/1611.00036}{{\ttfamily 1611.00036}}].

\bibitem{2017MNRAS.465.1757G}
H.~{Gil-Mar{\'\i}n}, W.~J. {Percival}, L.~{Verde}, J.~R. {Brownstein}, C.-H.
  {Chuang}, F.-S. {Kitaura} et~al., \emph{{The clustering of galaxies in the
  SDSS-III Baryon Oscillation Spectroscopic Survey: RSD measurement from the
  power spectrum and bispectrum of the DR12 BOSS galaxies}},
  \href{https://doi.org/10.1093/mnras/stw2679}{\emph{\mnras} {\bfseries 465}
  (2017) 1757} [\href{https://arxiv.org/abs/1606.00439}{{\ttfamily
  1606.00439}}].

\bibitem{2017arXiv170903452S}
E.~{Sellentin}, A.~H. {Jaffe} and A.~F. {Heavens}, \emph{{On the use of the
  Edgeworth expansion in cosmology I: how to foresee and evade its pitfalls}},
  {\emph{arXiv e-prints} (2017) arXiv:1709.03452}
  [\href{https://arxiv.org/abs/1709.03452}{{\ttfamily 1709.03452}}].

\bibitem{2021arXiv210208384P}
O.~H.~E. {Philcox} and Z.~{Slepian}, \emph{{Beyond Yamamoto: Anisotropic Power
  Spectra and Correlation Functions with Pairwise Lines-of-Sight}},
  {\emph{arXiv e-prints} (2021) arXiv:2102.08384}
  [\href{https://arxiv.org/abs/2102.08384}{{\ttfamily 2102.08384}}].

\bibitem{2020JCAP...05..042I}
M.~M. {Ivanov}, M.~{Simonovi{\'c}} and M.~{Zaldarriaga}, \emph{{Cosmological
  parameters from the BOSS galaxy power spectrum}},
  \href{https://doi.org/10.1088/1475-7516/2020/05/042}{\emph{\jcap} {\bfseries
  2020} (2020) 042} [\href{https://arxiv.org/abs/1909.05277}{{\ttfamily
  1909.05277}}].

\bibitem{2018AJ....156..160H}
N.~{Hand}, Y.~{Feng}, F.~{Beutler}, Y.~{Li}, C.~{Modi}, U.~{Seljak} et~al.,
  \emph{{nbodykit: An Open-source, Massively Parallel Toolkit for Large-scale
  Structure}}, \href{https://doi.org/10.3847/1538-3881/aadae0}{\emph{\aj}
  {\bfseries 156} (2018) 160}
  [\href{https://arxiv.org/abs/1712.05834}{{\ttfamily 1712.05834}}].

\bibitem{nist_dlmf}
{NIST}, \emph{NIST Digital Library of Mathematical Functions}. DLMF.

\bibitem{2000MNRAS.317..965H}
A.~F. {Heavens}, R.~{Jimenez} and O.~{Lahav}, \emph{{Massive lossless data
  compression and multiple parameter estimation from galaxy spectra}},
  \href{https://doi.org/10.1046/j.1365-8711.2000.03692.x}{\emph{\mnras}
  {\bfseries 317} (2000) 965}
  [\href{https://arxiv.org/abs/astro-ph/9911102}{{\ttfamily
  astro-ph/9911102}}].

\bibitem{2000ApJ...544..597S}
R.~{Scoccimarro}, \emph{{The Bispectrum: From Theory to Observations}},
  \href{https://doi.org/10.1086/317248}{\emph{\apj} {\bfseries 544} (2000) 597}
  [\href{https://arxiv.org/abs/astro-ph/0004086}{{\ttfamily
  astro-ph/0004086}}].

\bibitem{2018MNRAS.476L..60A}
J.~{Alsing} and B.~{Wandelt}, \emph{{Generalized massive optimal data
  compression}}, \href{https://doi.org/10.1093/mnrasl/sly029}{\emph{\mnras}
  {\bfseries 476} (2018) L60}
  [\href{https://arxiv.org/abs/1712.00012}{{\ttfamily 1712.00012}}].

\bibitem{2017MNRAS.470.2617A}
S.~{Alam}, M.~{Ata}, S.~{Bailey}, F.~{Beutler}, D.~{Bizyaev}, J.~A. {Blazek}
  et~al., \emph{{The clustering of galaxies in the completed SDSS-III Baryon
  Oscillation Spectroscopic Survey: cosmological analysis of the DR12 galaxy
  sample}}, \href{https://doi.org/10.1093/mnras/stx721}{\emph{\mnras}
  {\bfseries 470} (2017) 2617}
  [\href{https://arxiv.org/abs/1607.03155}{{\ttfamily 1607.03155}}].

\bibitem{2011AJ....142...72E}
D.~J. {Eisenstein}, D.~H. {Weinberg}, E.~{Agol}, H.~{Aihara}, C.~{Allende
  Prieto}, S.~F. {Anderson} et~al., \emph{{SDSS-III: Massive Spectroscopic
  Surveys of the Distant Universe, the Milky Way, and Extra-Solar Planetary
  Systems}}, \href{https://doi.org/10.1088/0004-6256/142/3/72}{\emph{\aj}
  {\bfseries 142} (2011) 72} [\href{https://arxiv.org/abs/1101.1529}{{\ttfamily
  1101.1529}}].

\bibitem{2016MNRAS.460.1173R}
S.~A. {Rodr{\'\i}guez-Torres}, C.-H. {Chuang}, F.~{Prada}, H.~{Guo},
  A.~{Klypin}, P.~{Behroozi} et~al., \emph{{The clustering of galaxies in the
  SDSS-III Baryon Oscillation Spectroscopic Survey: modelling the clustering
  and halo occupation distribution of BOSS CMASS galaxies in the Final Data
  Release}}, \href{https://doi.org/10.1093/mnras/stw1014}{\emph{\mnras}
  {\bfseries 460} (2016) 1173}
  [\href{https://arxiv.org/abs/1509.06404}{{\ttfamily 1509.06404}}].

\bibitem{2016MNRAS.456.4156K}
F.-S. {Kitaura}, S.~{Rodr{\'\i}guez-Torres}, C.-H. {Chuang}, C.~{Zhao},
  F.~{Prada}, H.~{Gil-Mar{\'\i}n} et~al., \emph{{The clustering of galaxies in
  the SDSS-III Baryon Oscillation Spectroscopic Survey: mock galaxy catalogues
  for the BOSS Final Data Release}},
  \href{https://doi.org/10.1093/mnras/stv2826}{\emph{\mnras} {\bfseries 456}
  (2016) 4156} [\href{https://arxiv.org/abs/1509.06400}{{\ttfamily
  1509.06400}}].

\bibitem{2020PhRvD.102f3533C}
A.~{Chudaykin}, M.~M. {Ivanov}, O.~H.~E. {Philcox} and M.~{Simonovi{\'c}},
  \emph{{Nonlinear perturbation theory extension of the Boltzmann code CLASS}},
  \href{https://doi.org/10.1103/PhysRevD.102.063533}{\emph{\prd} {\bfseries
  102} (2020) 063533} [\href{https://arxiv.org/abs/2004.10607}{{\ttfamily
  2004.10607}}].

\bibitem{2016arXiv160200674B}
T.~{Baldauf}, M.~{Mirbabayi}, M.~{Simonovi{\'c}} and M.~{Zaldarriaga},
  \emph{{LSS constraints with controlled theoretical uncertainties}},
  {\emph{arXiv e-prints} (2016) arXiv:1602.00674}
  [\href{https://arxiv.org/abs/1602.00674}{{\ttfamily 1602.00674}}].

\bibitem{2019JCAP...11..034C}
A.~{Chudaykin} and M.~M. {Ivanov}, \emph{{Measuring neutrino masses with
  large-scale structure: Euclid forecast with controlled theoretical error}},
  \href{https://doi.org/10.1088/1475-7516/2019/11/034}{\emph{\jcap} {\bfseries
  2019} (2019) 034} [\href{https://arxiv.org/abs/1907.06666}{{\ttfamily
  1907.06666}}].

\bibitem{2019PDU....24..260B}
T.~{Brinckmann} and J.~{Lesgourgues}, \emph{{MontePython 3: Boosted MCMC
  sampler and other features}},
  \href{https://doi.org/10.1016/j.dark.2018.100260}{\emph{Physics of the Dark
  Universe} {\bfseries 24} (2019) 100260}
  [\href{https://arxiv.org/abs/1804.07261}{{\ttfamily 1804.07261}}].

\bibitem{2007A&A...464..399H}
J.~{Hartlap}, P.~{Simon} and P.~{Schneider}, \emph{{Why your model parameter
  confidences might be too optimistic. Unbiased estimation of the inverse
  covariance matrix}},
  \href{https://doi.org/10.1051/0004-6361:20066170}{\emph{\aap} {\bfseries 464}
  (2007) 399} [\href{https://arxiv.org/abs/astro-ph/0608064}{{\ttfamily
  astro-ph/0608064}}].

\bibitem{2017MNRAS.464.4658S}
E.~{Sellentin} and A.~F. {Heavens}, \emph{{Quantifying lost information due to
  covariance matrix estimation in parameter inference}},
  \href{https://doi.org/10.1093/mnras/stw2697}{\emph{\mnras} {\bfseries 464}
  (2017) 4658} [\href{https://arxiv.org/abs/1609.00504}{{\ttfamily
  1609.00504}}].

\bibitem{2009arXiv0906.0664H}
A.~{Heavens}, \emph{{Statistical techniques in cosmology}}, {\emph{arXiv
  e-prints} (2009) arXiv:0906.0664}
  [\href{https://arxiv.org/abs/0906.0664}{{\ttfamily 0906.0664}}].

\bibitem{2005PhRvD..72d3003B}
D.~{Babich}, \emph{{Optimal estimation of non-Gaussianity}},
  \href{https://doi.org/10.1103/PhysRevD.72.043003}{\emph{\prd} {\bfseries 72}
  (2005) 043003} [\href{https://arxiv.org/abs/astro-ph/0503375}{{\ttfamily
  astro-ph/0503375}}].

\end{thebibliography}\endgroup

\end{document}